\documentclass{aa}  

\usepackage{graphicx}
\usepackage{txfonts}
\usepackage{placeins}
\usepackage{float}
\newcommand{\kms}{km\,s$^{-1}$}
\newcommand{\vr}{$V_{\rm r}$}

\newcommand{\prot}{$P_{\rm rot}$}

\newcommand{\vsini}{$v\sin i$}

\begin{document}

   \title{Characterising the magnetospheric accretion process of DF Tauri's primary}

   \author{K. Pouilly\inst{1}
          \and
          M. Audard\inst{1}
          }

   \institute{Department of Astronomy, University of Geneva, Chemin Pegasi 51, CH-1290 Versoix, Switzerland\\
              \email{Kim.Pouilly@unige.ch}
             }

   \date{Received 9 September 2025; accepted 21 February 2026}

 
  \abstract
   {The accretion process in young stellar objects (YSOs) is fundamental to the formation of stellar systems. This process governs the star's mass assembly, the transfer of angular momentum, and the shaping of the protoplanetary disc, thereby influencing planet formation.
   For classical T Tauri stars (cTTSs), which are low-mass YSOs, accretion is a well-understood process. Their strong, dipolar magnetic field truncates the disc at a few stellar radii. Material is then channelled along these magnetic field lines, creating accretion funnel flows that fall onto the star's surface.
   However, this paradigm, known as magnetospheric accretion, is limited to isolated stars. The accretion process in multiple systems has not yet been fully understood.}
   {This work is part of a series of studies designed to build a framework to understand the accretion process in multiple star systems. The specific goal here is to determine how the magnetospheric accretion model can be used to describe \object{DF Tau}, a binary system where only the primary star is accreting material.}
   {To investigate how accretion occurs in a system where a single star is orbited by a non-accreting stellar companion, we used a time series of high-resolution spectropolarimetric observations from the ESPaDOnS instrument. This allowed us to study the accretion-related emission line variability, the veiling, and the magnetic field topology of the primary star in the system.}
   {Our research concludes that the primary star of the DF Tau system undergoes typical magnetospheric accretion. This process is driven by a strong dipolar magnetic field, which funnels accreting material onto the stellar surface, creating an accretion shock.
    We also identified a significant difference in the magnetic topology of the two stars querying the influence of accretion of the evolution of the magnetic field, or capture of the secondary star.}
   {}

   \keywords{Stars: variables: T Tauri --
                Stars: individual: DF Tau --
                Stars: magnetic field --
                Accretion, accretion disks --
                Techniques: spectroscopic --
                Techniques: polarimetric
               }

   \maketitle
%

\section{Introduction}
\label{sec:intro}
    The accretion process of pre-main sequence (PMS) stars is a cornerstone of stellar and planetary formation.
    It is a fundamental mechanism that governs the transfer of angular momentum between a star and its surrounding disc, a crucial factor in the star's stability and development.
    Furthermore, accretion plays a vital role in shaping the structure of the protoplanetary disc itself, which is the birthplace of exoplanets. 
    Consequently, a comprehensive understanding of the accretion process is a major objective in the study of how entire stellar systems form and evolve over time.

    Classical T Tauri stars (cTTSs) are low-mass PMS stars that are surrounded by an accretion disc.
    These objects possess a powerful dipole magnetic field that acts to truncate the inner edge of the disc at a distance of just a few stellar radii.
    The disc material is then channelled along the magnetic field lines directly onto the star's surface, forming the so-called accretion funnel flows, producing an accretion shock \citep[see the review by][]{Hartmann16}.
    This widely observed process is known as magnetospheric accretion and is the dominant accretion paradigm for the majority of cTTSs.

    However, this traditional scheme is based on the assumption that a star is isolated and surrounded by a single circumstellar disc.
    This assumption is challenged by the fact that most stars are born in multiple systems \citep{Offner23}.
    As a result, our understanding of the accretion mechanisms in young stellar objects that are part of binary or multiple systems remains incomplete and requires further investigation.
    The present study of the young binary system \object{DF Tau} is part of our ongoing research to bridge this significant knowledge gap \citep[see our other recent studies of DQ Tau, AK Sco, V4046 Sgr, and EX Lup by][respectively]{Pouilly23, Pouilly24c, Pouilly24, Pouilly25}.

    DF Tau is a young binary system, consisting of two similar-mass M2 PMS stars with a 48-year orbital period, assumed as coeval \citep{Allen17}.
    The projected orbital separation is approximately 100 mas, which translates to a physical distance of 14 au, given the distance of DF Tau of 142 pc \citep{Krolikowski21, Gaia23}.
    Interestingly, while both components were initially classified as cTTSs by \cite{Hartigan03}, \cite{Allen17} deduced from Keck/NIRSPEC component spectra that the primary star is the sole member of the pair currently undergoing active material accretion. The authors thus assumed that this component only has an accretion disc, meaning that the disc signature detected by \cite{Hartigan03} on the secondary is the result of a contamination by the primary's disc.
    The authors also derived stellar parameters for the two components, resulting on a \vsini\ of 13 \kms\ for the primary, and 41 \kms\ for the secondary, consistent with the shorter rotation period they derived for the secondary (\prot\ = 10.5 and 3.33 days for the primary and secondary, respectively). They also concluded that the inclination of the rotation axis of both components is about 90$^\circ$ based on the relation between the rotation periods and the \vsini\ of the two stars.

    Recent analysis by \cite{Kutra25} re-examined Keck/NIRSPEC component spectra initially studied by \cite{Allen17}, and added ALMA observation, providing new insights into the system's properties.
    The authors found that both components have a disc, with inclination of about 41 and 46$^\circ$, which are reasonably well aligned with the orbital inclination of 54.3$^\circ$, but concluded that only the primary has an inner disc. 
    Regarding the individual stellar parameters, they found the primary to have a $T_{\rm eff}$ of 3638 K, a \vsini\ of 16.4 \kms, a mean small-scale magnetic field of 2.5 kG, and a significant veiling at 1560 nm ($r$ = 1.4). The veiling is a reduction in the depth of absorption lines due to excess continuum emission by the accretion shock \citep{Hartigan91}. It is thus a typical signature of accretion onto the star. In contrast, the secondary star has a similar $T_{\rm eff}$ (3433 K) and small-scale magnetic field (2.6 kG) but rotates much more rapidly (\vsini = 46.2 \kms) and shows either a very weak or entirely absent veiling.

    The primary objective of this work is to characterise the accretion process of the DF~Tau primary. 
    We will examine its detailed interaction with the stellar magnetic field and investigate how this process conforms to the established magnetospheric accretion paradigm for cTTSs, particularly in the context of a binary system.

    The remainder of this paper is structured as follows: in Section \ref{sec:obs}, we provide a comprehensive description of the high-resolution spectropolarimetric observations that were used in this study.
    Section \ref{sec:results} presents our results, which are then thoroughly discussed in Section \ref{sec:discussion}. 
    Finally, we present our conclusions of this work in Section \ref{sec:conclusion}.
    
\section{Observations}
\label{sec:obs}
    
    The time series used in this work was acquired using the Echelle SpectroPolarimetric Device for the Observation of Stars (ESPaDOnS) \citep{Donati03}, which is mounted on the 3.6 m telescope at the Canada-France-Hawaii Telescope (CFHT). 
    All observations were conducted in spectropolarimetric mode.

    Each observation thus consists of four sub-exposures taken in different polarimeter configurations.
    These were then combined to produce the intensity (unpolarised, Stokes \textit{I}), the circularly polarised (Stokes \textit{V}), and the null polarisation (Stokes \textit{N}) spectra of DF Tau.
    Each observation was reduced using the CFHT in-built reduction package for ESPaDOnS \texttt{Libre-ESpRIT} \citep{Donati97} and are available on the \texttt{PolarBase} database \citep{Donati97, Petit14}.

    A log of observation is provided in Table~\ref{tab:obs}.
    \begin{table}
    \centering
    \caption{Log of ESPaDOnS observations of DF Tau.}
    \begin{tabular}{lllll}
        \hline
        Date & HJD & S/N$_{\rm I}$ & S/N$_{\rm LSD}$ & \vr \\
        (2011) & ($-$2\,450\,000 d) & & & (\kms) \\
        \hline
        01 Nov & 5866.98 & 113 & 2223 & 16.96$\pm$0.28 \\
        05 Nov & 5870.99 & 77 & 1211 & 18.15$\pm$0.06 \\
        06 Nov & 5871.96 & 31 & 714 &  15.68$\pm$0.52 \\
        07 Nov & 5873.01 & 77 & 1431 & 17.85$\pm$0.13 \\
        09 Nov & 5874.93 & 171 & 1718 & 19.92$\pm$0.16 \\
        11 Nov & 5876.97 & 154 & 2059 & 17.31$\pm$0.30 \\
        13 Nov & 5878.94 & 164 & 2097 & 13.20$\pm$0.23 \\
        14 Nov & 5879.90 & 173 & 2675 & 20.16$\pm$0.05 \\
        15 Nov & 5880.97 & 172 & 2990 & 14.99$\pm$0.06 \\
        16 Nov & 5882.07 & 166 & 2734 & 17.79$\pm$0.16 \\
        \hline  
    \end{tabular}
    \tablefoot{S/N$_{\rm I}$ is the peak S/N by spectral pixel at order 31 (730 nm), S/N$_{\rm LSD}$ corresponds to the effective S/N of the LSD Stokes \textit{I} profiles (see Sect.~\ref{subsec:lsd}), and \vr\ correspond to the radial velocity of the primary computed from the 2D CCF (see Sect.~\ref{subsec:vrad}).}
    \label{tab:obs}
    \end{table}

\section{Results}
\label{sec:results}

    \subsection{Radial velocities}
    \label{subsec:vrad}

    The first step of our analysis was to determine the radial velocity (\vr) of the primary component of the system for each observation. We employed a cross-correlation method using a synthetic spectrum across several wavelength windows. This synthetic spectrum was computed using the \texttt{ZEEMAN} code \citep{Landstreet88, Wade01, Folsom12} with a line list from the \texttt{VALD} database \citep{Ryabchikova15} and \texttt{MARCS} atmospheric models \citep{Gustafsson08}. We used the stellar parameters for the primary component as determined by \cite{Kutra25}. A Gaussian fit was then applied to the resulting cross-correlation function (CCF) to derive the \vr\ values.

    However, the ESPaDOnS spectra obtained at the CFHT were not able to resolve the two components of the system. Given the small amplitude of the primary's orbital \vr\ curve and the high rotational velocity of the secondary, we hypothesised that the secondary's contribution would appear as a slightly shifted, broad component. This would distort the primary's photospheric profile and therefore falsify the \vr\ values obtained from a one-dimensional (1D) CCF.

    To investigate this hypothesis, we conducted a two-dimensional (2D) CCF analysis using two synthetic spectra, which were computed with the parameters estimated by \cite{Kutra25}. We then fitted the 2D CCF with a two-dimensional Gaussian to obtain the velocities of both the primary and secondary components. The \vr\ values obtained for the primary were significantly higher than those from the 1D CCF and were far more consistent with the orbital \vr\ solution provided by \cite{Allen17}.
    More details are provided in Appendix~\ref{ap:vr}.

    While the velocity of the primary is more accurately recovered using the 2D CCF, the determination of the secondary's broad and shallow component velocity is highly dependent on the S/N of the observation in a given wavelength window, leading to a significant uncertainty in the obtained values. However, a precise estimate of the secondary's \vr\ is outside the scope of this paper. The radial velocities for the primary of DF Tau are summarised in Table \ref{tab:obs}.

    \subsection{Emission lines}
    \label{subsec:emlines}

    Since the primary star is the only component in the system that is accreting, the emission lines discussed in this section are attributed to it. These lines are crucial for understanding the overall accretion pattern of the system.
    We analysed five distinct emission lines that probe the structure of the accretion flow and shock of a cTTS.
    The Balmer lines, specifically H$\alpha$, H$\beta$, and H$\gamma$, are partially formed within the accretion funnel's flow itself \citep{Muzerolle01}.
    Next, we studied the \ion{Ca}{ii} Infrared Triplet (IRT). The narrow component (NC) of this triplet is known to form in an extended chromosphere, close to the accretion shock \citep{Donati11}.
    Finally, we analysed the \ion{He}{i} D$_{3}$ line. The NC of this line is produced directly within the accretion shock \citep{Beristain01}, offering a view into the innermost region of the accretion process.

    \subsubsection{Balmer lines}
    \label{subsubsec:balmer}
    
\begin{figure*}
        \centering
        \includegraphics[width=0.38\linewidth]{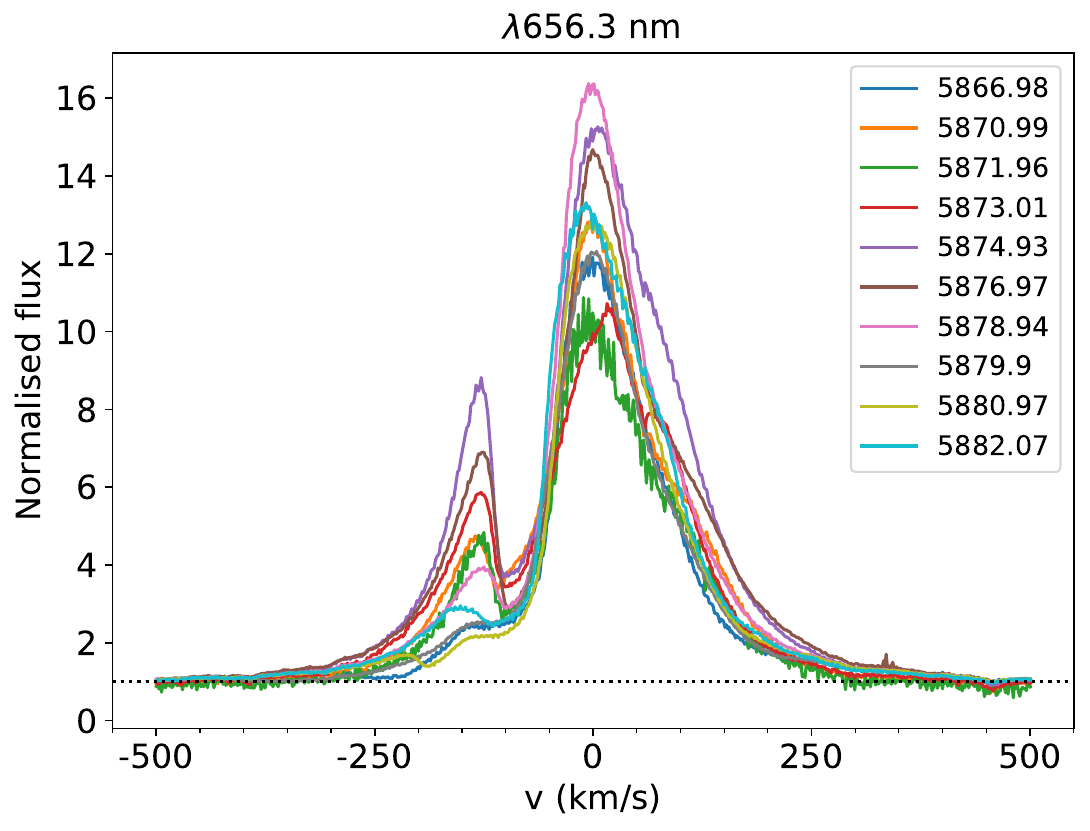}
        \includegraphics[width=0.3\linewidth]{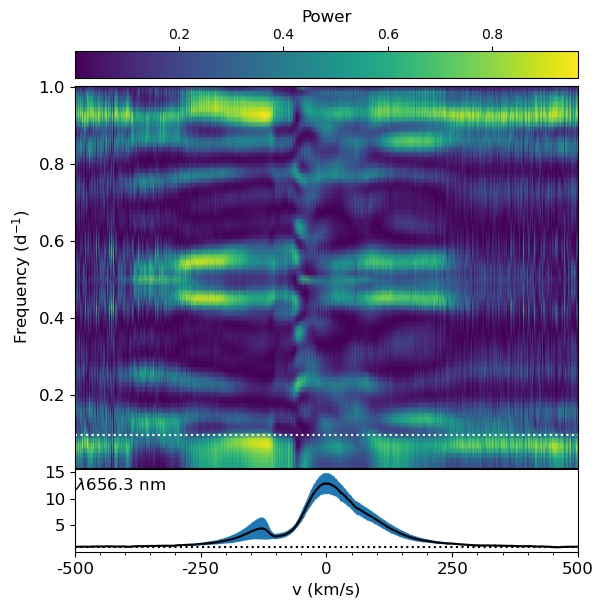}
        \includegraphics[width=0.3\linewidth]{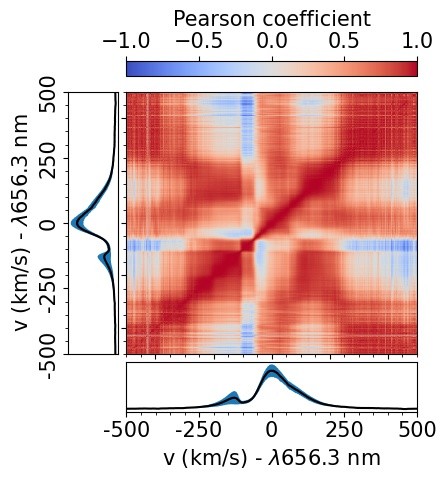}
 
        \includegraphics[width=0.38\linewidth]{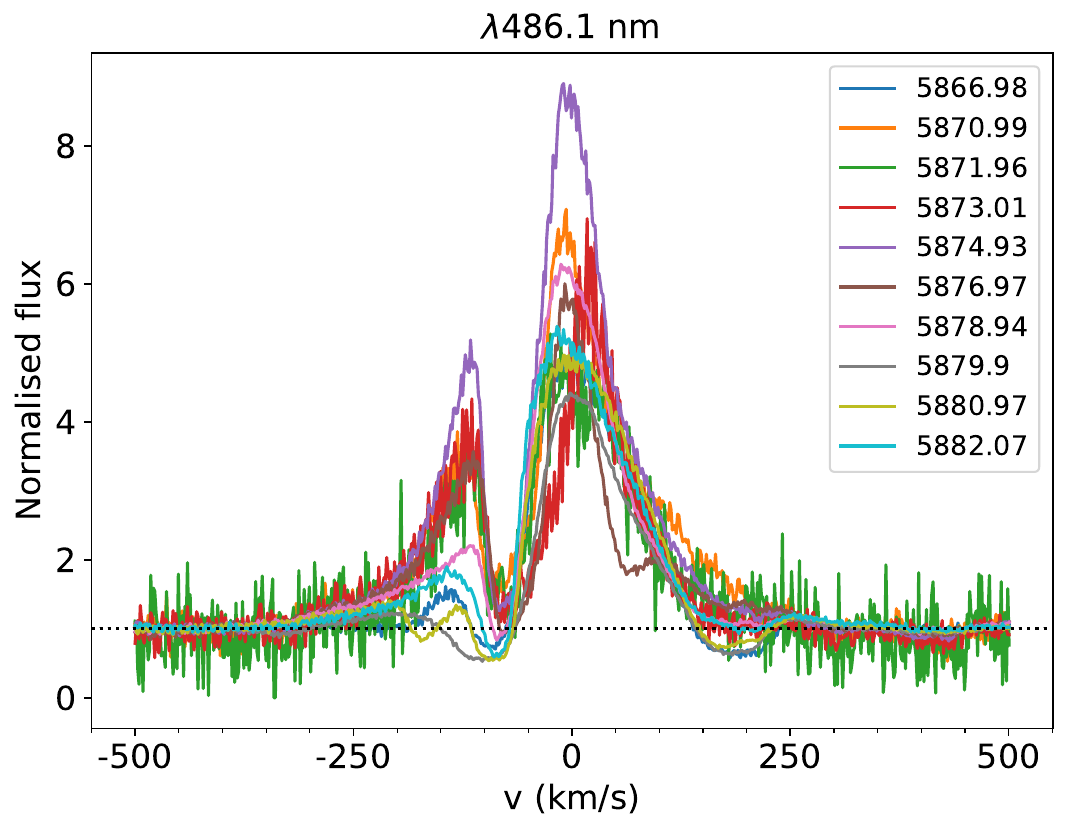}
        \includegraphics[width=0.3\linewidth]{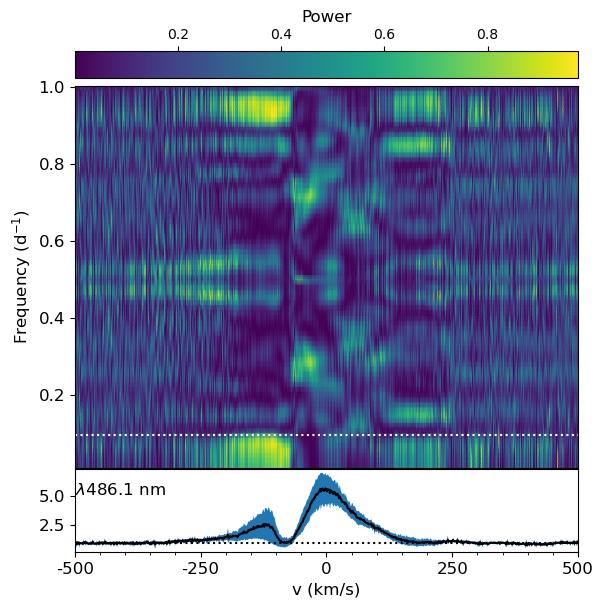}
        \includegraphics[width=0.3\linewidth]{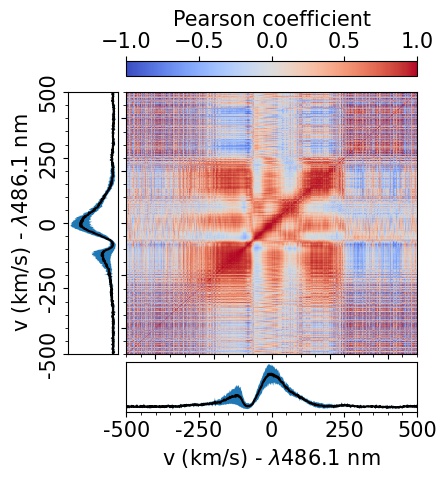}
   
        \includegraphics[width=0.38\linewidth]{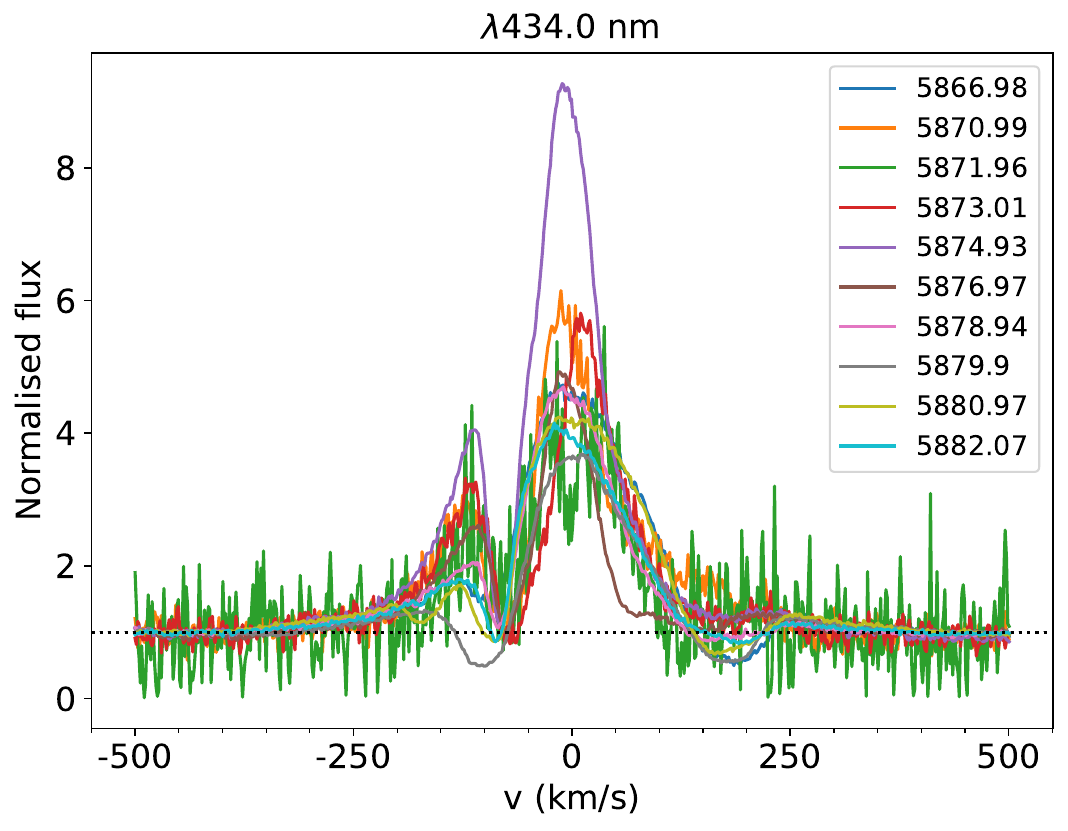}
        \includegraphics[width=0.3\linewidth]{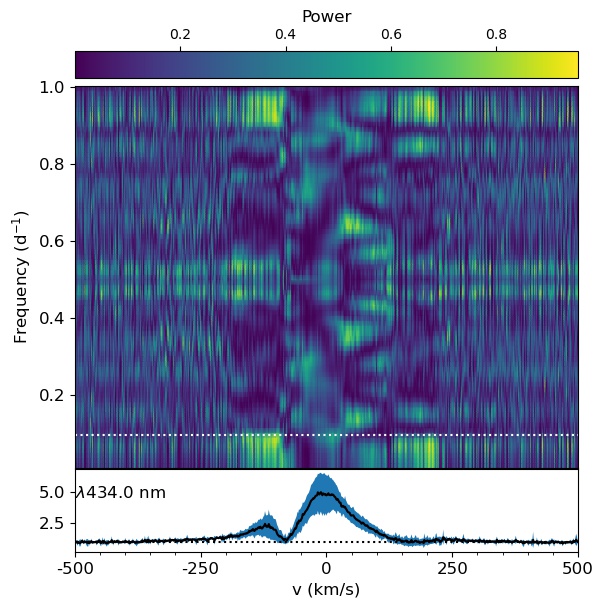}
        \includegraphics[width=0.3\linewidth]{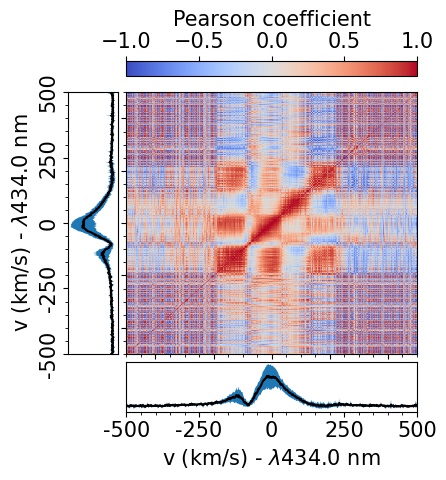}
   
        \caption{Variability analysis of Balmer lines. The H$\alpha$, H$\beta$, and H$\gamma$ lines are depicted on the first, second, and third line, respectively. The first column shows the line profiles, each colour corresponding to an observation. The second column are the P2Ds, with the velocity on the x-axis, the frequency on the y-axis, and the power of the periodogram scaled by the colour bar. The white dotted line shows the primary rotation period of 10.5 days derived by \cite{Allen17}. The mean profile and its variance are shown in black and blue, respectively, along the x-axis of the P2D. The third column present the ACMs. The two lines velocity are on the x and y axes and the colour bar scaled the Pearson correlation coefficient, red being 1, highly correlated, by being $-$1, highly anticorrelated. The two mean and variance profiles are illustrated in black and blue, respectively, along its corresponding axis. }
        \label{fig:balmer}
    \end{figure*}

    The conditions within the accretion funnel flow itself are ideal to promote the emission of the Balmer lines (H$\alpha$, H$\beta$, and H$\gamma$) \citep{Muzerolle01}. As shown in Fig.~\ref{fig:balmer}, the H$\alpha$, H$\beta$, and H$\gamma$ lines, corrected for the primary's velocity, all display very similar double-peaked profiles and strong variability.

    The H$\beta$ and H$\gamma$ lines show larger variability than the H$\alpha$ line, notably a stronger peak at HJD 2\,455\,874.93 with respect to other observations such as at HJD 2\,455\,878.94. They also exhibit significant blue- and red-shifted absorption that dips below the continuum level. The red-shifted absorption is particularly noteworthy as it is a characteristic sign of accreting material moving into the observer's line of sight. This specific type of spectral signature is known as an Inverse P-Cygni (IPC) profile.

    To investigate the periodicity of this variability, we computed 2-dimensional periodograms (P2D). A P2D consists of a generalised Lomb-Scargle (GLS) periodogram calculated for each velocity channel of the spectral line.

    The P2Ds reveal that the blue- and red-shifted absorption features are modulated on a period slightly longer than the star's reported 10.5-day rotation period, with False Alarm Probability \citep[FAP, computed from the prescription of][]{Baluev08} reaching 2$\times$10$^{-3}$, 8$\times$10$^{-4}$, and 1$\times$10$^{-2}$ for H$\alpha$, H$\beta$, and H$\gamma$, respectively. A FAP lower than 10$^{-1}$ is considered as statistically significant for the P2D in this work. This type of accretion-related variability, synchronised with the stellar rotation period, is an expected behaviour within the magnetospheric accretion paradigm.

    To break down the different sources of variability within the spectral lines, we computed correlation matrices (CMs), which show the linear correlation coefficient between the velocity channels of two lines. A strong correlation highlights that a variability is dominated by a single physical process. Strong anti-correlations also indicate linked processes. The auto-correlation matrices (ACMs), which are CMs calculated between a line and itself, are presented for the H$\alpha$, H$\beta$, and H$\gamma$ lines in Fig.~\ref{fig:balmer}.

    These matrices reveal four main correlated regions: the blue-shifted emission, the central emission, the low-velocity red-shifted absorption, and the IPC profile.
    On can note the absence of correlation, nor anticorrelation, between the various regions. This means that there are four physical processes are in place in these lines, which we interpret as follow the central emission is simply the line emission. The IPC profile region's variability is dominated by the passage of the accretion column only. The blue shifted regions probably translate the drastic flux depletion observed that can be the effect of a wind. Finally, the low velocity red-shifted regions can be a low-velocity wind at the foot of the accretion funnel flow, seen as redshifted due geometrical effect as previously observed on AA Tau and V807 Tau \citep[][respectively]{Bouvier03, Pouilly21}.
    
    \subsubsection{\ion{Ca}{ii} IRT}
    \label{subsubsec:ca}

\begin{figure}
        \centering
        \includegraphics[width=0.54\linewidth]{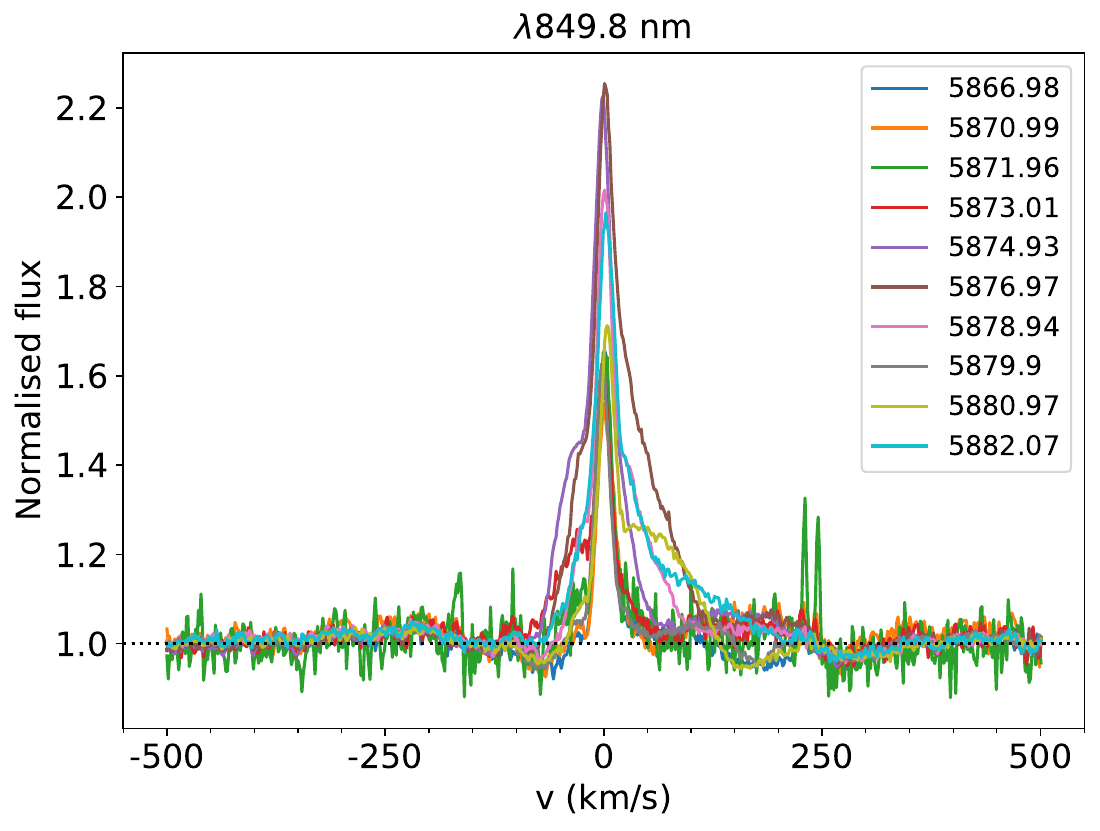}
        \includegraphics[width=0.45\linewidth]{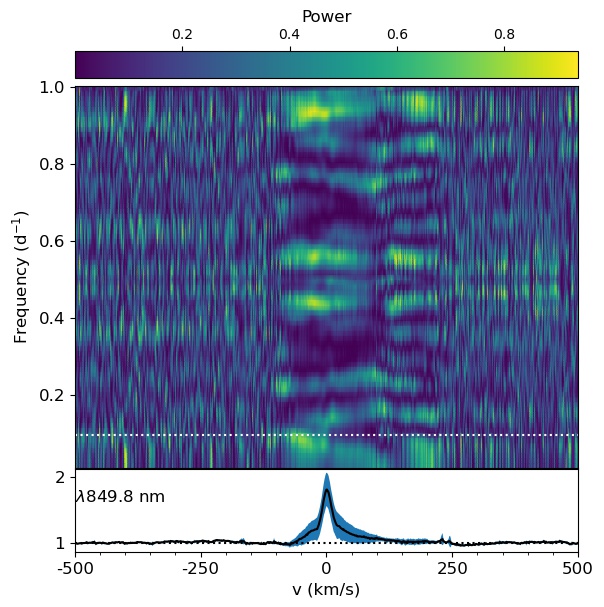}
        \includegraphics[width=0.49\linewidth]{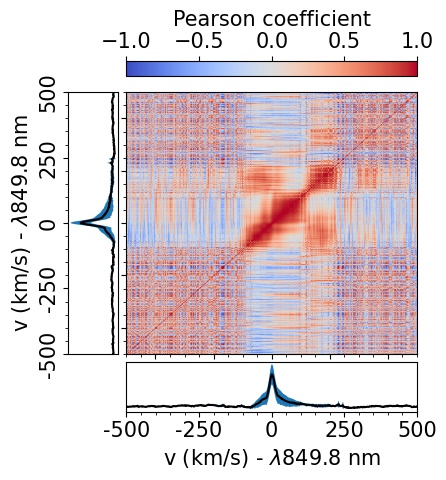}
        \includegraphics[width=0.49\linewidth]{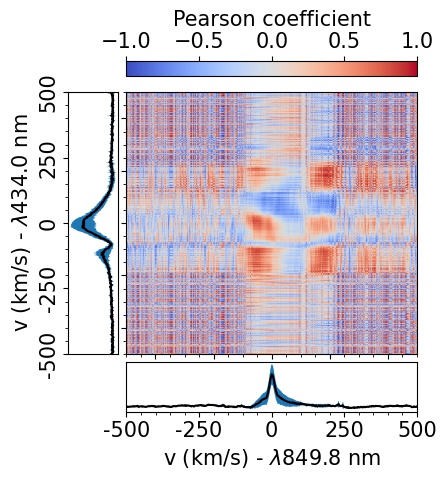}
        \caption{Variability analysis of the \ion{Ca}{ii} IRT line. The top left panel illustrates the line profiles, the top right panel shows the P2D, the bottom left panel present the ACM, and the bottom right panel depicts the CM with H$\gamma$. The colours and figure shapes are the same as used in Fig.~\ref{fig:balmer}. }
        \label{fig:caIIirt}
    \end{figure}

    The next lines we studied are the \ion{Ca}{ii} IRT. Because of the similar shape and variability of the three lines of the triplet, we focused on the 849.8-nm component and referred it as \ion{Ca}{ii} IRT for simplicity. Its NC is known to form in an extended chromosphere, near the accretion shock. As shown in Figure \ref{fig:caIIirt}, the DF Tau's \ion{Ca}{ii} IRT line is made up of a highly variable NC superimposed on a highly variable broad component (BC). It also shows blue- and red-shifted absorption, with the latter being consistent with the IPC profiles seen in the Balmer lines.

    The P2D (see Fig.~\ref{fig:caIIirt}) reveals periodic behaviour. Specifically, the blue-shifted absorption shows a periodicity consistent with the stellar rotation period, but has to be taken carefully due to its high FAP (10$^{-1}$). The IPC profile region also displays a periodicity, but on a slightly longer period (FAP=10$^{-2}$), which aligns with the period detected in the Balmer lines.

    The ACM, also in Fig.~\ref{fig:caIIirt}, highlights three distinct correlated regions: a blue-shifted region, a centred region, and a red-shifted region. When correlated with the H$\gamma$ line, the CM (Figure \ref{fig:caIIirt}) indicates only a slight anti-correlation ($-$0.7) between the NC and the IPC profile of H$\gamma$.

    This anti-correlation is expected if the accretion shock and the accretion column are aligned, as the former process induces a peak in emission while the latter causes absorption. The low value of the anti-correlation likely indicates a slight phase shift between the passage of these two features into the observer's line of sight.
    
    \subsubsection{\ion{He}{i} D$_3$}
    \label{subsubsec:he}

    The final emission line analysed in this study is the \ion{He}{i} D$_3$ line. Its NC is a well-known feature in accretion studies because it is formed in the post-shock region of the accretion shock, located on the star's surface \citep{Beristain01}.

\begin{figure}
        \centering
        \includegraphics[width=0.54\linewidth]{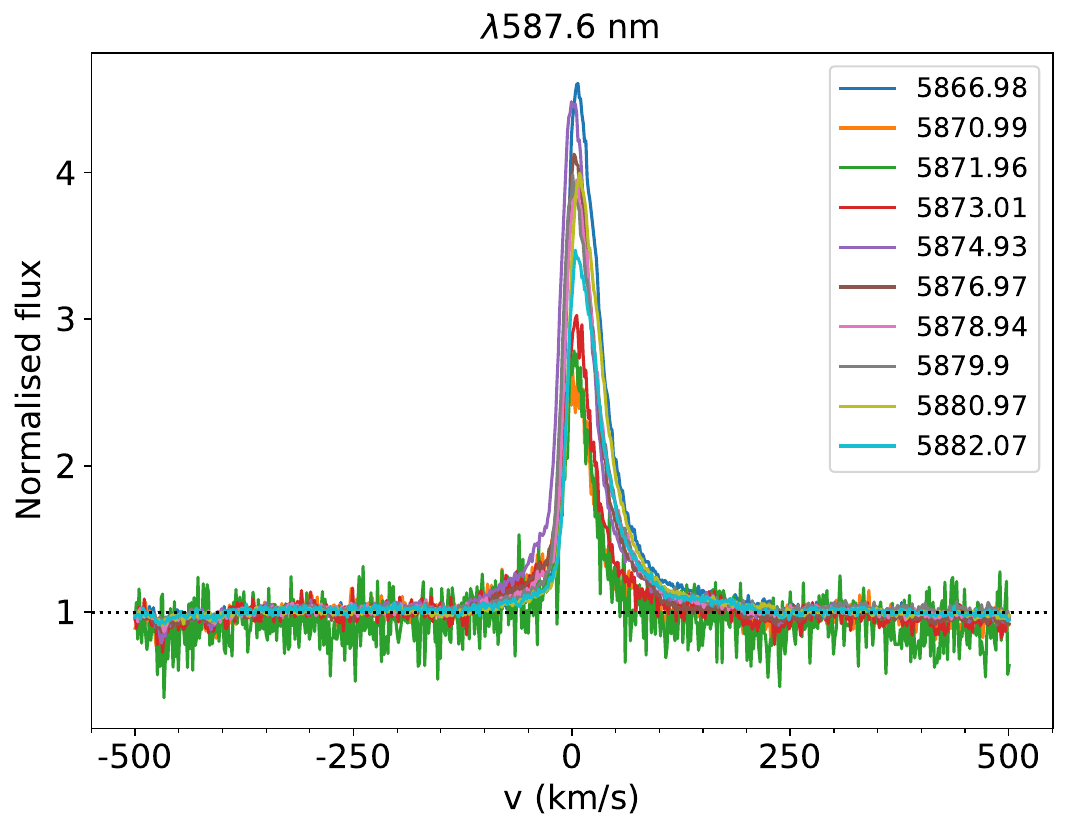}
        \includegraphics[width=0.45\linewidth]{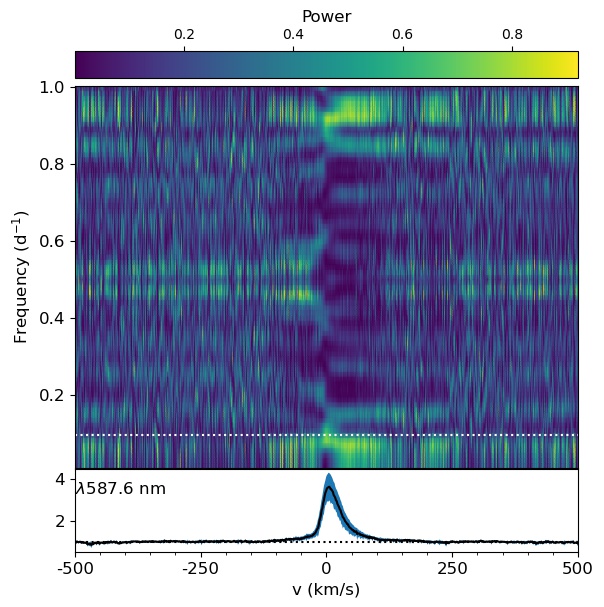}
        
        \includegraphics[width=0.49\linewidth]{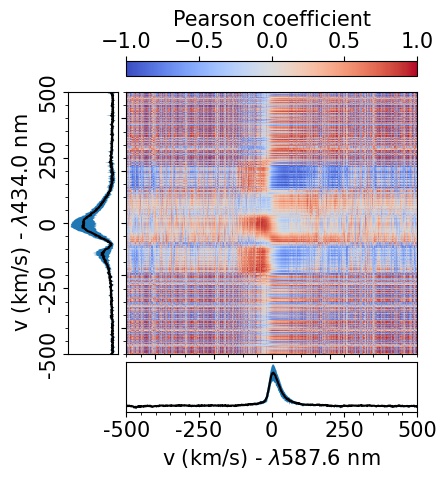}
        \includegraphics[width=0.49\linewidth]{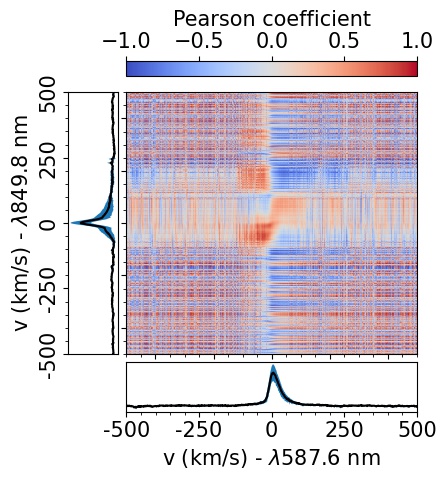}
        \caption{Variability analysis of the \ion{He}{i} D$_{3}$ line. The top left panel illustrates the line profiles, the top right panel shows the P2D, the bottom left panel present the CM with the \ion{Ca}{ii} IRT, and the bottom right panel depicts the CM with H$\gamma$. The colours and figure shapes are the same as used in Fig.~\ref{fig:balmer}. }
        \label{fig:heId3}
    \end{figure}

    As shown in Fig.~\ref{fig:heId3}, the \ion{He}{i} D$_3$ line in DF Tau consists of a very strong NC and a weak BC. The NC exhibits a modulation period that is consistent with the modulation of the IPC profiles observed in the Balmer and \ion{Ca}{ii} IRT lines, with a FAP of 3$\times$10$^{-2}$.

    The CM between the \ion{He}{i} D$_3$ line and the H$\gamma$ line (also in Fig.~\ref{fig:heId3}) reveals a slight anti-correlation ($-$0.7) between the \ion{He}{i} D$_3$ NC and the H$\gamma$ IPC profile region. When cross-correlated with the \ion{Ca}{ii} IRT, only a slight correlation between the NCs is recovered. This finding confirms that the two emission regions are close but are not fully aligned with each other, nor are they aligned with the accretion funnel flow.

    \subsection{Least square deconvolution profiles and veiling}
    \label{subsec:lsd}

    To further our analysis, we computed the least-squares deconvolution (LSD) profiles \citep{Donati97, Kochukhov10} of the ESPaDOnS spectra. This technique consist of a weighted average of as many photospheric lines as possible, in this case, about 12\,000 lines. The lines are selected using a line mask based on the \texttt{VALD} database line list \citep{Ryabchikova15}. This process yields both unpolarised (Stokes \textit{I}) and circularly polarised (Stokes \textit{V}) LSD profiles. To normalise the LSD weights used for the weighted average aforementioned, we used mean line strength, Landé factor, and wavelength values of 0.2, 1.3, and 500 nm, respectively.

    Since DF Tau is a spectroscopic binary, both components are present in the profiles. The raw LSD profiles, shown in Appendix~\ref{ap:rawLSD}, have a S/N ranging from 2990 to 714. Fortunately, the secondary star rotates much faster than the primary, resulting in significantly broader profiles. This difference allows us to correct for the secondary's contribution in Stokes \textit{I} by fitting a double Voigt profile and assuming a luminosity ratio ($LR$) of 1, given that the two components have very similar masses, spectral types, and presumably coeval \citep{Allen17}. A example of this decomposition is provided in Appendix~\ref{ap:rawLSD}. A Stokes \textit{V} signature is detected in all profiles at the primary's velocity range, but none is found at the secondary's velocity, so no correction is needed for Stokes \textit{V}.

    In Fig.~\ref{fig:rawlsd}, we can observe the large variability in the amplitude of the Stokes \textit{I} profiles. While the secondary component appears stable, the depth of the primary component shows significant variation. We attribute this to substantial veiling. Such a large variability must be quantified and corrected for to allow for a proper analysis of the profiles.

    To estimate the veiling, we used a combined spectrum from two photospheric templates. The template used was TWA7, a weak-lined T Tauri star (meaning it is no longer accreting) with $T_{\rm eff}$ = 3800 K, \vr\ of 13.18 \kms, and a projected rotational velocity (\vsini) of 4.5 \kms \citep{Nichoslon21}. This template was corrected from its \vr, rotationally broadened to match the \vsini\ of the primary (16.4 \kms) and secondary (46.2 \kms), and assumed an $LR$ of 1.

\begin{figure}
        \centering
        \includegraphics[width=0.9\linewidth]{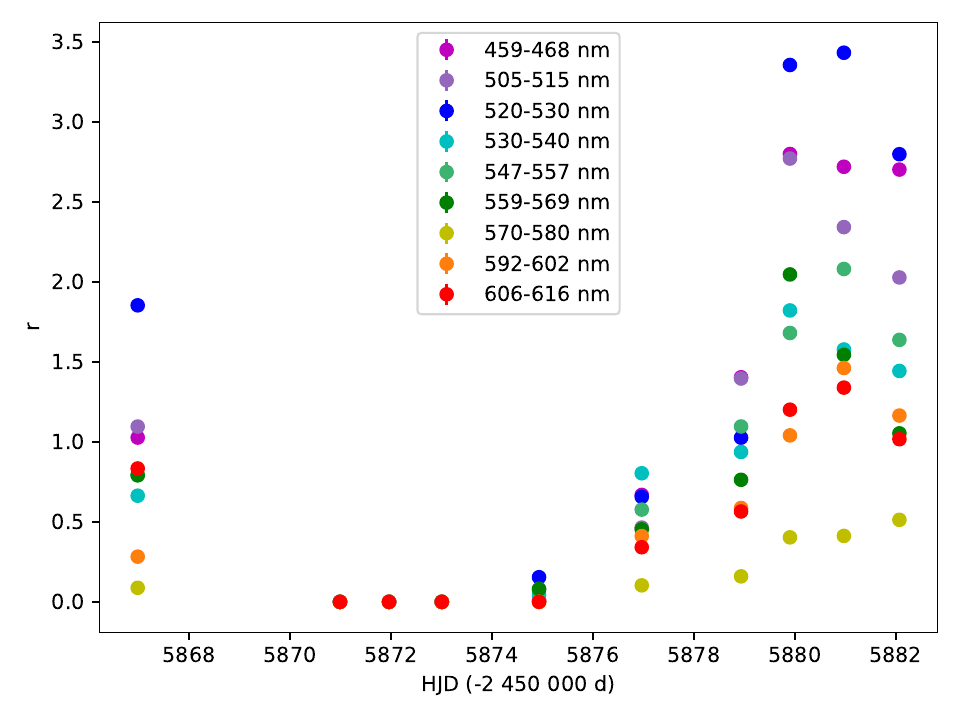}
        \caption{Veiling values as function of the HJD. The colours indicate the different wavelength windows used.}
        \label{fig:veil}
    \end{figure}

    We then fitted the veiling on the primary component using a Levenberg-Marquardt algorithm (LMA) with the following equation:

    \begin{equation}
        I = \frac{I_0 + r}{1+r},
    \end{equation}

    \noindent where $r$ is the veiling, $I$ is the veiled spectrum, and $I_0$ is the unveiled spectrum. This fit was performed on nine wavelength windows ranging from 460 to 616 nm. We carefully selected these windows to be well-normalised, contain numerous photospheric lines, and be free of emission or heavily blended lines. Since the secondary is not accreting, we assumed an absence of veiling for this component. The results of this analysis are shown in Fig.~\ref{fig:veil}.

    We can see that the veiling is indeed highly variable, both across different observations and within different wavelength windows. To correct the Stokes \textit{I} profiles for this effect, we used the average veiling value calculated across all the windows. These corrected profiles are shown in Fig.~~\ref{fig:lsdfit}, alongside the Stokes \textit{V} profiles.

    \subsection{Magnetic topology}
    \label{subsec:mag}
    
\begin{figure*}
        \centering
        \includegraphics[width=0.45\linewidth]{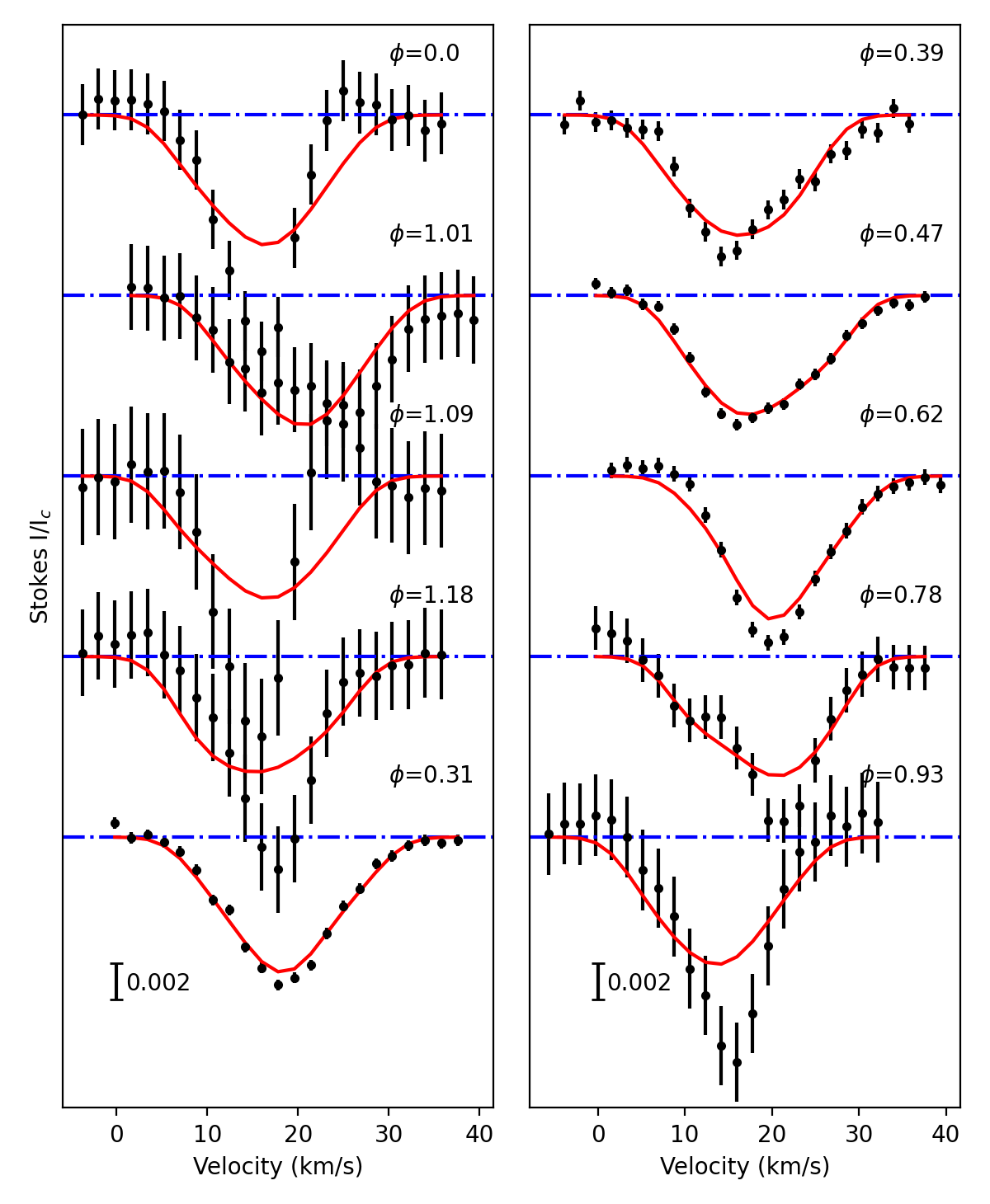}
        \includegraphics[width=0.45\linewidth]{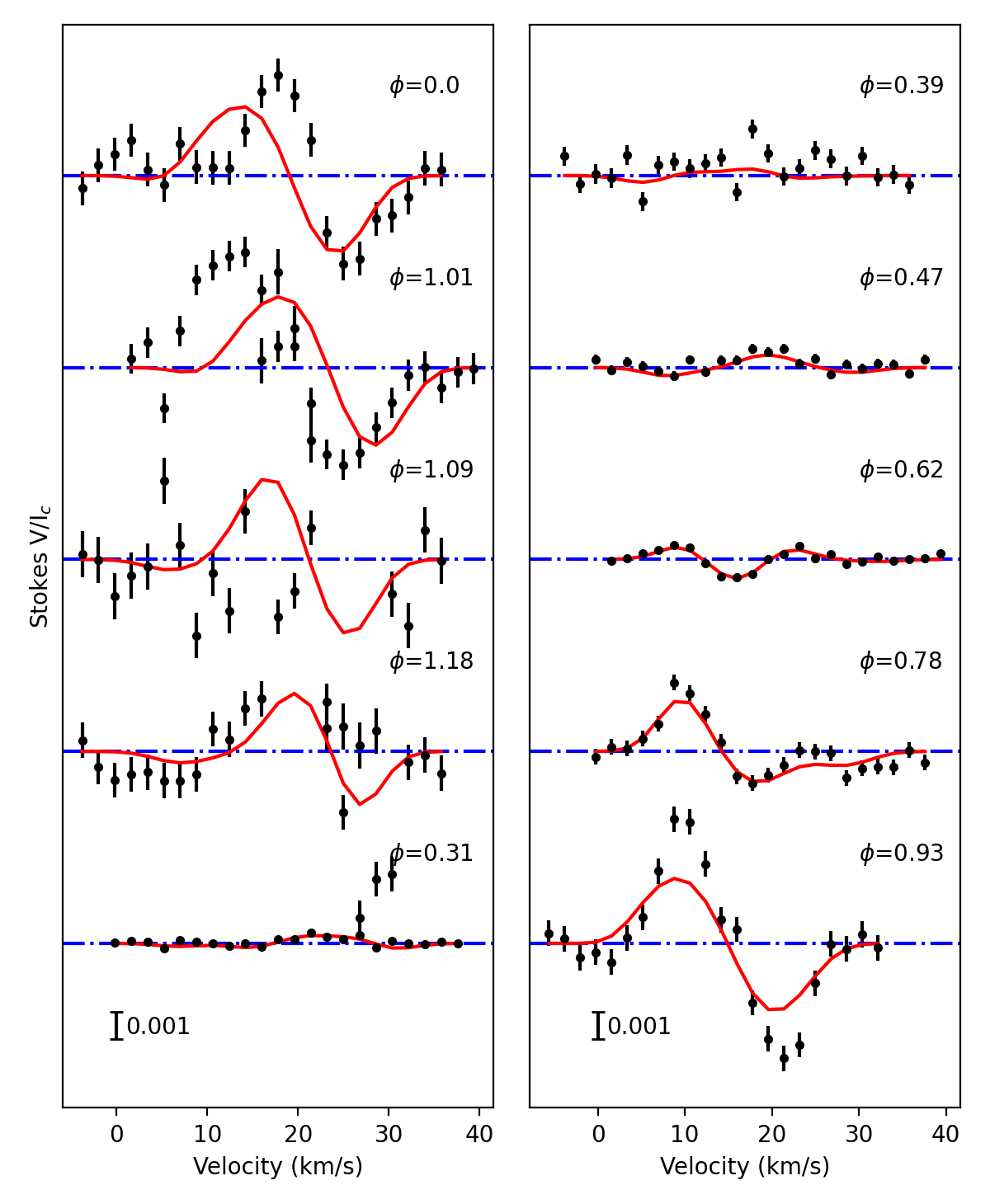}
        \caption{LSD Stokes \textit{I} (\textit{left}) corrected from the secondary's contribution and Stokes \textit{V} (\textit{right}) profiles (\textit{black dots}) and their fit using the ZDI analysis (\textit{red}). The profiles are ordered by phase, the latter being indicated on the right of each profile. To see the profile with the contribution of the secondary, please refer to Fig.~\ref{fig:rawlsd}.}
        \label{fig:lsdfit}
    \end{figure*}

    Finally, we used the LSD profiles computed in Sect.~\ref{subsec:lsd} and shown in Fig.~\ref{fig:lsdfit} to conduct a complete Zeeman-Doppler Imaging (ZDI) analysis on the primary only.
    Indeed, the secondary does not show any Stokes \textit{V} signature (see Sect.~\ref{subsec:lsd}) despite the similar small-scale field than the primary derived by \cite{Kutra25}.
    This is probably due to a more complex magnetic topology that lower the contribution of the large-scale magnetic field on the line of sight. We performed this using the \texttt{ZDIpy} implementation by \cite{Folsom18}.

    First, we reconstructed the Doppler image (DI) using the Stokes \textit{I} profiles. This procedure begins with a uniform brightness distribution and then iteratively adds bright and dark features to match the entire Stokes \textit{I} dataset. The local Voigt profile parameters were estimated by fitting the LSD profile of the TWA7 template.

\begin{figure}
        \centering
        \includegraphics[width=0.9\linewidth]{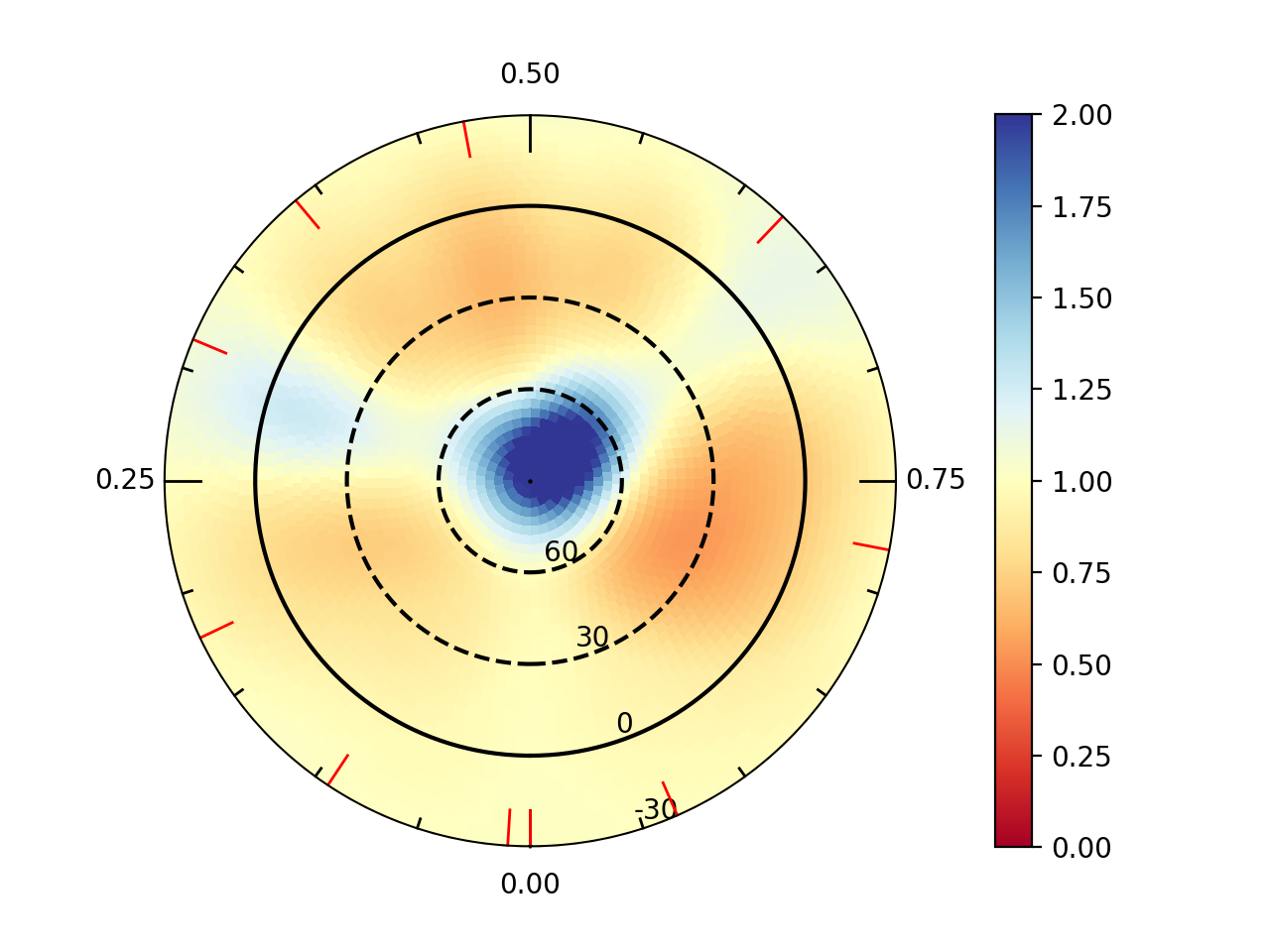}
        \caption{Doppler image (brightness maps) of DF Tau on a flattened polar view. The central dot illustrate the rotation pole, the two dotted circles are latitude 60 and 30$^\circ$, and the solid circle represents the equator. The black ticks show the clockwise rotation phases, and the red ticks represent the observed phases. The colour-code indicates the brightness on a linear scale, where a value of 1 represents the quiet photosphere. Values lower than one are darker regions, and values greater than one are bright.}
        \label{fig:di}
    \end{figure}

    This analysis requires precise estimates of several stellar atmospheric parameters, including the rotation period (\prot), the inclination ($i$), and the \vsini. We therefore used ZDI to optimise these values by computing the DI on a grid of parameters and selecting the minimal $\chi^2$ to find the best-fit values. This process resulted in a \prot\ of 12.8 days, an $i$ of 54.6$^\circ$, and a \vsini\ of 12.0 \kms. The \vsini\ is fully consistent with the value derived by \cite{Allen17} and the rotation period is slightly higher than the 10.5 days they obtained, which is reminiscent of the emission lines variability (see Sect.~\ref{subsec:emlines}). The inclination we derived is not consistent with the 90$^\circ$ assumed by the authors, but is consistent with the disc and orbit inclination \cite[40--55$^\circ$,][]{Kutra25}.
    The resulting DI is shown in Fig.~\ref{fig:di} and reveals a bright polar feature that extends up to 45$^\circ$ latitude towards phase 0.62.

\begin{figure}
        \centering
        \includegraphics[width=0.9\linewidth]{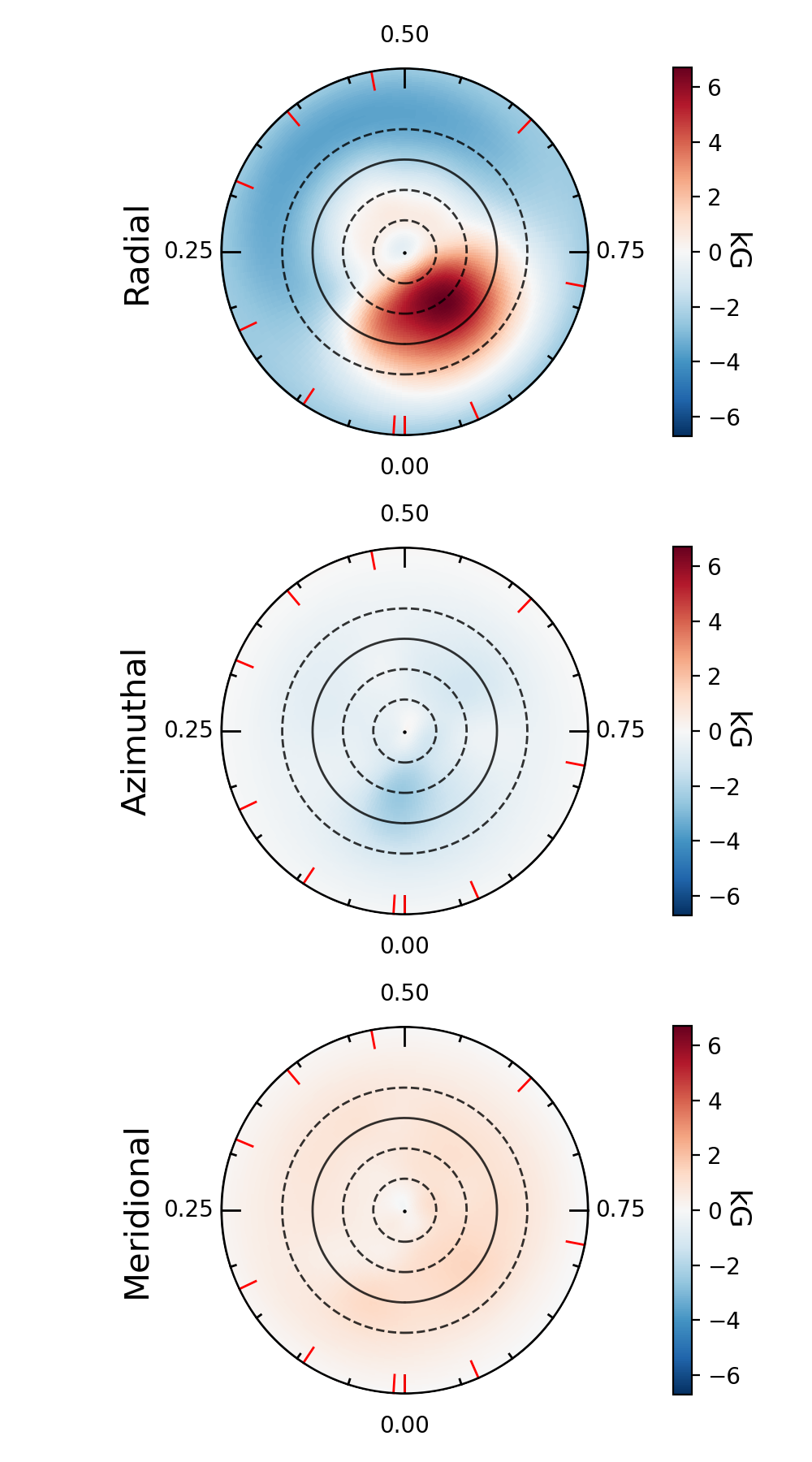}
        \caption{Magnetic maps of DF Tau. The radial, azimuthal, and meridional fields are illustrated on the top, middle, and bottom panels, respectively. We used the same flattened polar view as Fig.~\ref{fig:di}. The colour-code scales the magnetic field strength from dark blue for the strongest negative value to dark red for the strongest positive value}
        \label{fig:magmaps}
    \end{figure}

    The magnetic topology was derived by adjusting its spherical harmonic components \citep{Donati06} using the Stokes \textit{V} profiles.
    The analysis revealed a magnetic topology that is overwhelmingly dominated by a poloidal field (92\%), with the dipole component being the most significant contributor (83\%). A noteworthy contribution from the quadrupole (13\%) was also identified.
    The dipole negative pole itself reaches a strength of $-$4 kG and is located at 34$^\circ$ latitude and phase 0.6. This position is in full agreement with the bright feature observed in the DI. The complete magnetic topology is summarised in Table \ref{tab:magtop}, and the corresponding magnetic maps are presented in Fig.~\ref{fig:magmaps}. The fit to the resulting profiles is shown in Fig.~\ref{fig:lsdfit}.

    \begin{table}
        \centering
        \caption{Magnetic topology of DF Tau}

        \begin{tabular}{lc}
            \hline
            \hline
             Poloidal field (\% of the total energy) &  92.2 \\
             Toroidal field (\% of the total energy) & 7.8 \\
             Dipole (\% of the poloidal field) & 82.8 \\
             Quadrupole (\% of the poloidal field) & 12.9 \\
             Octupole (\% of the poloidal field) & 3.0 \\
             Axisymmetry (\% of the poloidal field) & 35.5 \\
             $\langle B \rangle$ (kG) & 2.5 \\
             $B_{\rm dip}$ (kG) & 4.2 \\
             \hline
        \end{tabular}
        \label{tab:magtop}
    \end{table}

\section{Discussion}
\label{sec:discussion}

    This work aimed to characterise the accretion process of the binary star DF Tau, as part of a broader study of accretion in binary systems \citep{Pouilly23, Pouilly24, Pouilly24c, Pouilly25}. We selected DF Tau due to its unique configuration: a binary system where only the primary star possesses a circumstellar disc and is actively accreting, while the non-accreting secondary acts as a gravitational perturber.

    Our analysis revealed that the primary star of DF Tau exhibits a typical magnetospheric accretion process. We found an accretion flow passing through the observer's line of sight, which creates an IPC profile on the Balmer and \ion{Ca}{ii} IRT lines. This flow's periodic nature aligns with the star's rotation period.

    The flow itself is driven by a strong, dipole-dominant magnetic topology with a field strength of $-$4 kG. The accretion shock, which is responsible for the emission in the NC of the \ion{He}{i} line \citep{Beristain01}, is located near the rotation pole, which is consistent with the position of the star's magnetic dipole pole.

\begin{figure}
        \centering
        \includegraphics[width=0.9\linewidth]{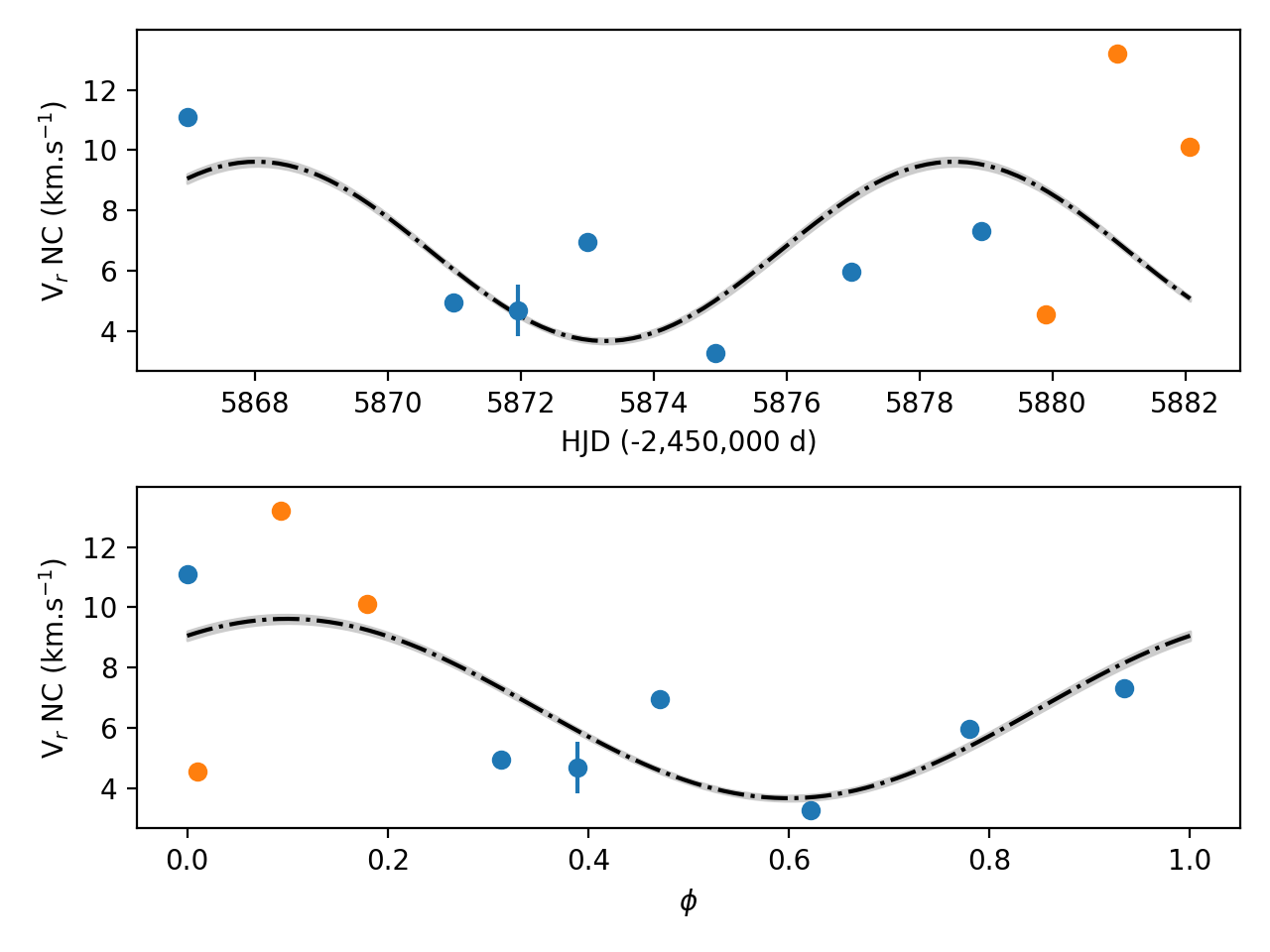}
    \caption{Radial velocity fit of NC of the \ion{He}{i} D$_3$ line (\textit{top}) and its version folded in phase (\textit{bottom}) using \prot=12.8 d (see Sect.~\ref{subsec:mag}). The colours indicate different rotation cycles.}
        \label{fig:vrhe}
    \end{figure}

    To confirm this, we used the same methodology as in our previous work \citep{Pouilly21, Pouilly24c, Pouilly25}. We recovered the location of the NC's emission by fitting a geometrical model to its velocity modulation, using the rotation period (\prot) of 12.8 days derived in Section \ref{subsec:mag}. This yielded an emission location at phase 0.95 and a latitude of 83$^\circ$, which is in complete agreement with the dipole pole's location. The results of this fit are shown in Figure \ref{fig:vrhe}.

    The shape of the Balmer lines in DF Tau is similar to that of other cTTSs known to exhibit magnetospheric accretion, such as AA Tau \citep{Bouvier03} and V807 Tau \citep{Pouilly21}. In these objects, the line shape is interpreted as a broad central emission with two narrow absorption features, and the correlation between their velocities can indicate magnetospheric inflation, a phenomenon characteristic of magnetospheric accretion. 

    It seems that the secondary component does not affect the magnetospheric accretion observed on the primary. This sets a limit on the separation and eccentricity for which the secondary component has to be considered when studying the accretion process.
    
    The rotation period we derived (12.8 days) is different from the 10.5 days reported by \cite{Allen17}, but it enabled a correct fit of the LSD profiles during our ZDI reconstruction. This 12.8-day period is also consistent with the periodic modulation we observed in the accretion-tracing emission lines (see Section \ref{subsec:emlines}).

    Another key difference is the derived inclination of the rotation axis. While \cite{Allen17} estimated a near edge-on inclination (90$^\circ$), our findings show an inclination of 54.6$^\circ$. This value is consistent with the hypothesis from \cite{Kutra25} that the rotation axis is aligned with the disc and orbital axes (40–55$^\circ$).

    One surprising result is the complete absence of a large-scale magnetic field detected on the secondary star, despite the similar small-scale field strengths reported for both components by \cite{Kutra25}. The simplest explanation for this is a more complex magnetic field topology on the secondary. 
    Indeed, the circular polarisation used to measure the magnetic field of the star is sensitive to the large-scale field on the line of sight only. A more complex magnetic topology may induce directly a lower contribution on the line of sight, and/or magnetic polarity cancellation at larger scale, resulting in the absence of Stokes \textit{V} signature despite the presence of a magnetic field.
    This might suggest that the lack of an inner disc and ongoing accretion has accelerated the "complexification" of the magnetic field, a process typically observed in slightly more evolved stars. This finding points to a two-way interaction: the magnetic field influences accretion, and the accretion process, in turn, influences the magnetic field, as suggested by X-ray observation of V1118~Ori \citep{Audard05, Audard10}.
    
    An alternative explanation could be that the secondary star was captured by the primary's system. If the two components are confirmed to be young PMS stars \citep[1--2 Myrs,][]{Herbig88, Chen90, Schaefer14}, they are only supposed by \cite{Allen17} to be coeval. This means that the two components might be at a different evolutionary stage, and that the secondary has stared the development of a radiative core inducing a complexification of the large-scale magnetic field topology.


\section{Conclusions}
\label{sec:conclusion}

    This paper summarises a study on the accretion process of the primary star in the DF Tau binary system. The primary is the only accreting component in the equal-mass, 100-mas-separated pair. The research aims to understand how a stellar-mass companion affects the standard magnetospheric accretion process. This work is part of a larger effort to investigate accretion in multiple systems \citep[see][for the study of DQ Tau, AK Sco, EX Lup, and V4046 Sgr, respectively.]{Pouilly23, Pouilly24, Pouilly24c, Pouilly25}.

    The study confirms that the primary star's accretion is magnetically driven, consistent with the magnetospheric accretion paradigm. A kG-strong dipole magnetic field truncates the circumstellar disc, channelling material along field lines into a free-falling accretion funnel. This funnel produces an accretion shock on the stellar surface near the dipole pole.

    The primary's rotation period was refined from a previous value of 10.5 days \citep{Allen17} to 12.8 days. The inclination of the rotation axis was also constrained to 54.6$^\circ$, a parameter that was previously not well-defined.

    A surprising finding was the absence of a large-scale magnetic field signature in Stokes \textit{V} for the secondary star, despite it having a similar small-scale field strength to the primary. This suggests a much more complex magnetic topology for the secondary. This finding raises two possible explanations: either the lack of accretion has accelerated the evolution of the secondary's magnetic field, or the secondary may have been captured by the primary's system.


\begin{acknowledgements}
    I (KP) would like to warmly thanks J. Bouvier and E. Alecian for their teaching and support during my career, nothing would have been possible without their incredible knowledge transmission.
    I would also like to thanks O. Kochukhov, A. Halhin and the whole Astronomy department of Uppsala University, as well as M. Audard and the whole team of Observatoire de Geneve for believing in my projects and myself, allowing me to live this wonderful journey in the Astrophysics community.
    
    This research was funded in whole or in part by the Swiss National Science Foundation (SNSF), grant number 217195 (SIMBA).
    
    Based on observations obtained at the Canada–France– Hawaii Telescope (CFHT) which is operated from the summit of Maunakea by the National Research Council of Canada, the institut National des Sciences de l’Univers of the Centre National de la Recherche Scientifique of France, and the University of Hawaii. The observations at the Canada–France–Hawaii Telescope were performed with care and respect from the summit of Maunakea which is a significant cultural and historic site.

    This work has made use of the VALD database, operated at Uppsala University, the Institute of Astronomy RAS in Moscow, and the University of Vienna.

    The \texttt{PySTEL(L)A} package is available at \url{https://github.com/pouillyk/PySTELLA}.

\end{acknowledgements}

\bibliographystyle{aa}
\bibliography{literature}

\begin{appendix}

\begin{onecolumn}
    \section{Radial velocity derivation}
    \label{ap:vr}
    In this appendix we present examples of CFF results to derive the radial velocity using two models: primary plus secondary and primary only.
    These are presented in Fig.~\ref{fig:ccf2d} and \ref{fig:ccf1d}, respectively.
    Figure \ref{fig:comp1d2d} shows the two corresponding radial velocity curves. One can note that the 2D CCF yield lower uncertainties, as well as more consistent results toward the velocities obtained from the component resolved spectra studied in \cite{Kutra25}.

    \begin{figure*}[!h]
        \centering
        \includegraphics[width=.98\linewidth]{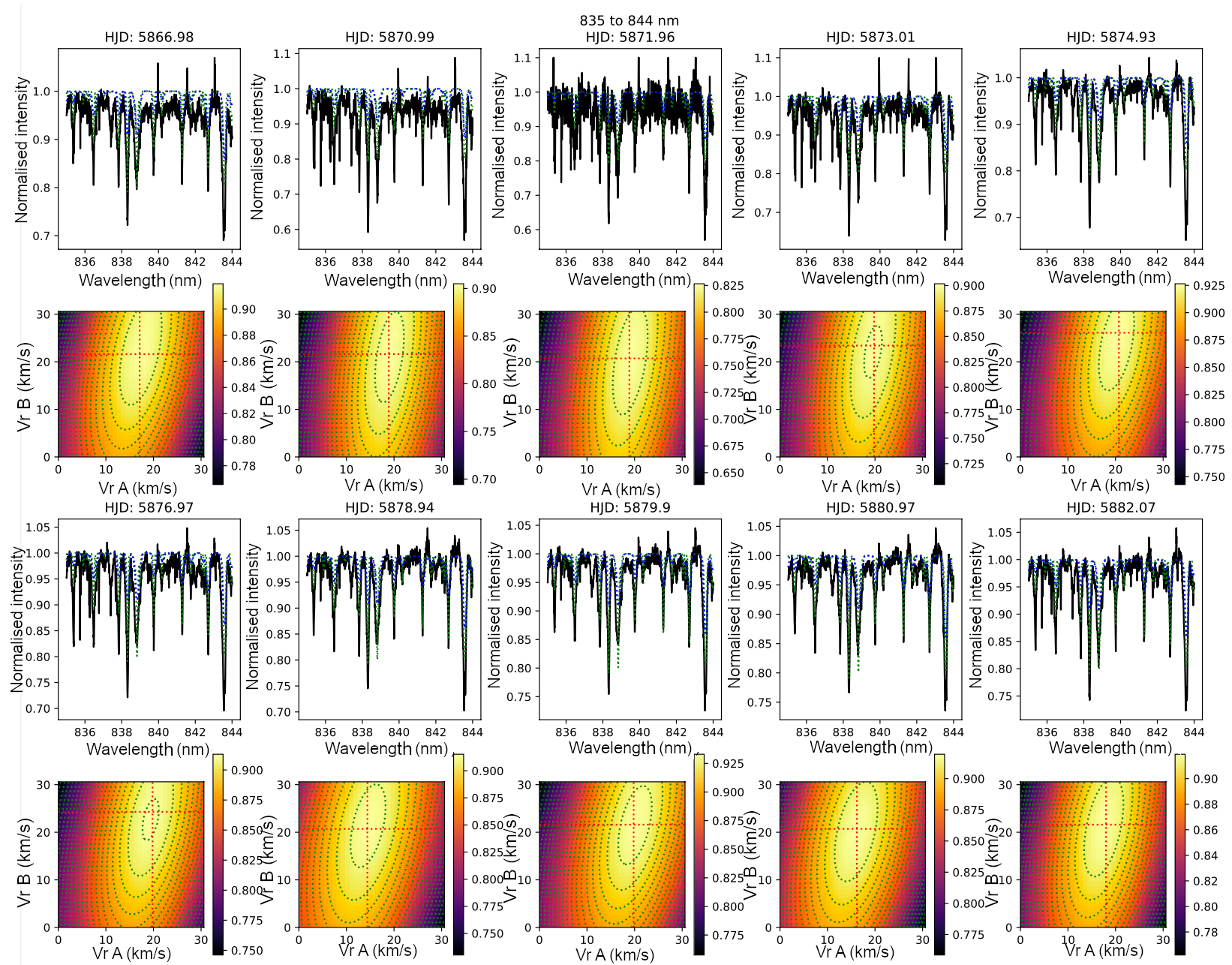}
        \caption{Cross correlation results on the 835--844 nm wavelength window for the primary plus secondary model. \textit{First and third rows:} Observed spectra (\textit{black}), template for the primary (\textit{blue}) and for the secondary (\textit{green}), shifted to the obtained radial velocities. \textit{Second and fourth rows :} 2D CCF obtained. The x- and y-axes represent the radial velocity of the primary and secondary, respectively. The colour scales is the Pearson correlation coefficient, the dotted green lines are the 2D Gaussian fit, and the red dotted line the peak of the fitted Gaussian.  }
        \label{fig:ccf2d}
    \end{figure*}

    \begin{figure*}
        \centering
        \includegraphics[width=0.49\linewidth]{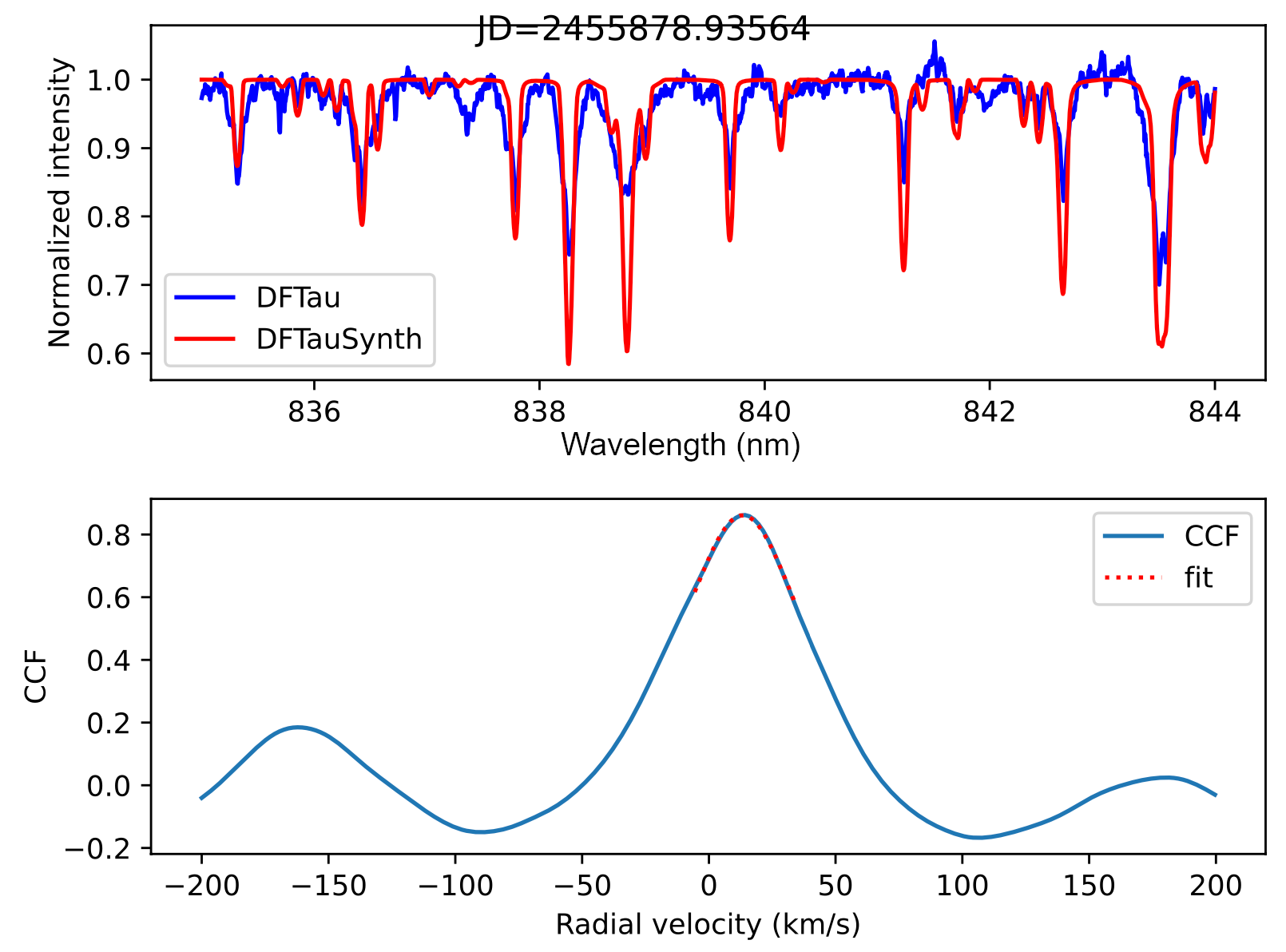}
        \includegraphics[width=0.49\linewidth]{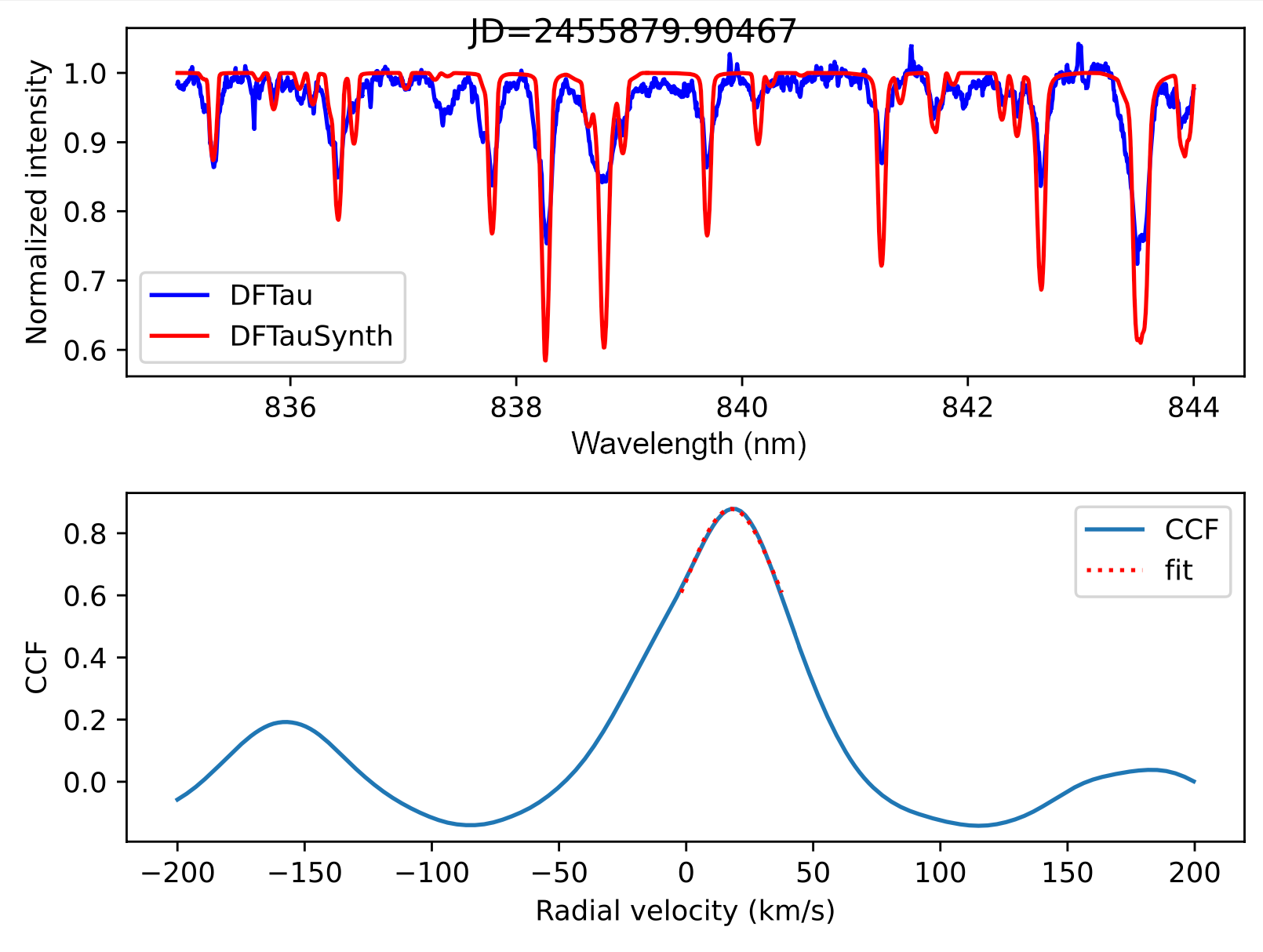}
        \includegraphics[width=0.49\linewidth]{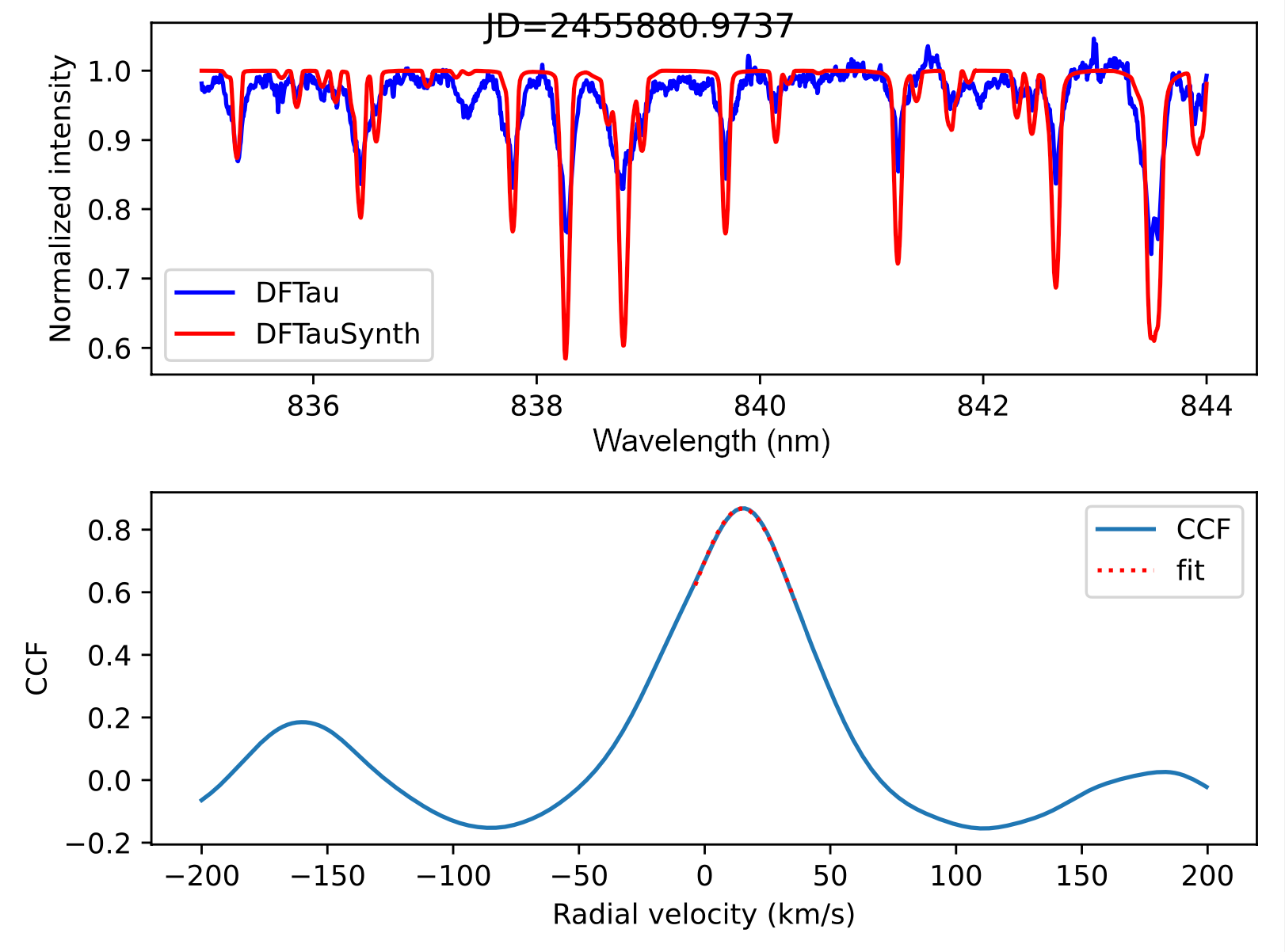}
        \includegraphics[width=0.49\linewidth]{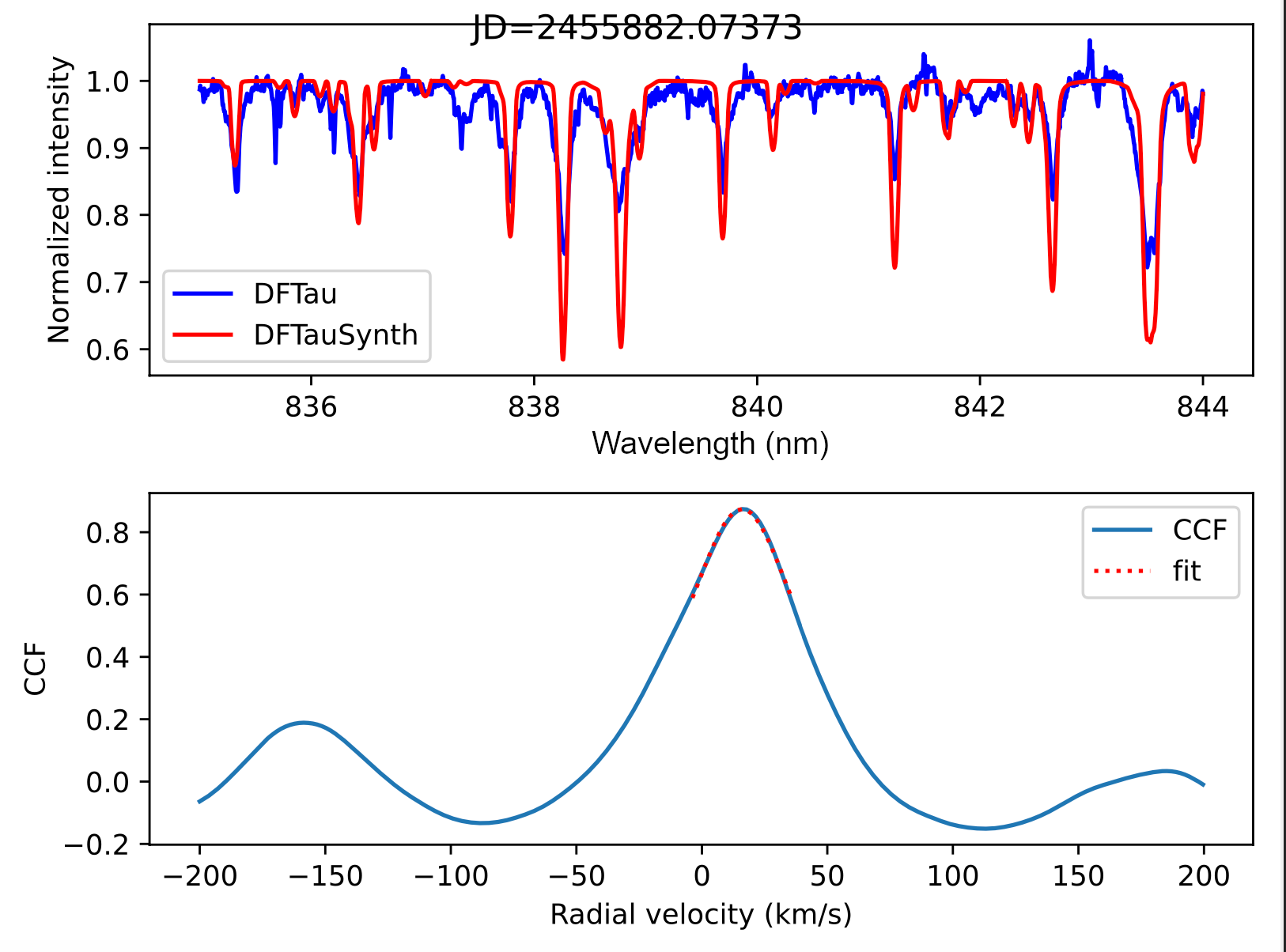}
        \caption{Cross correlation results on the 835--844 nm wavelength window for the last four observation using the primary only model. \textit{First and third rows:} Observed spectra (\textit{blue}) and template (\textit{red}) shifted to the obtained radial velocity. \textit{Second and fourth rows :} CCF (\textit{blue}) and Gaussian fit \textit{(red}). }
        \label{fig:ccf1d}
    \end{figure*}

    \begin{figure}
        \centering
        \includegraphics[width=0.70\linewidth]{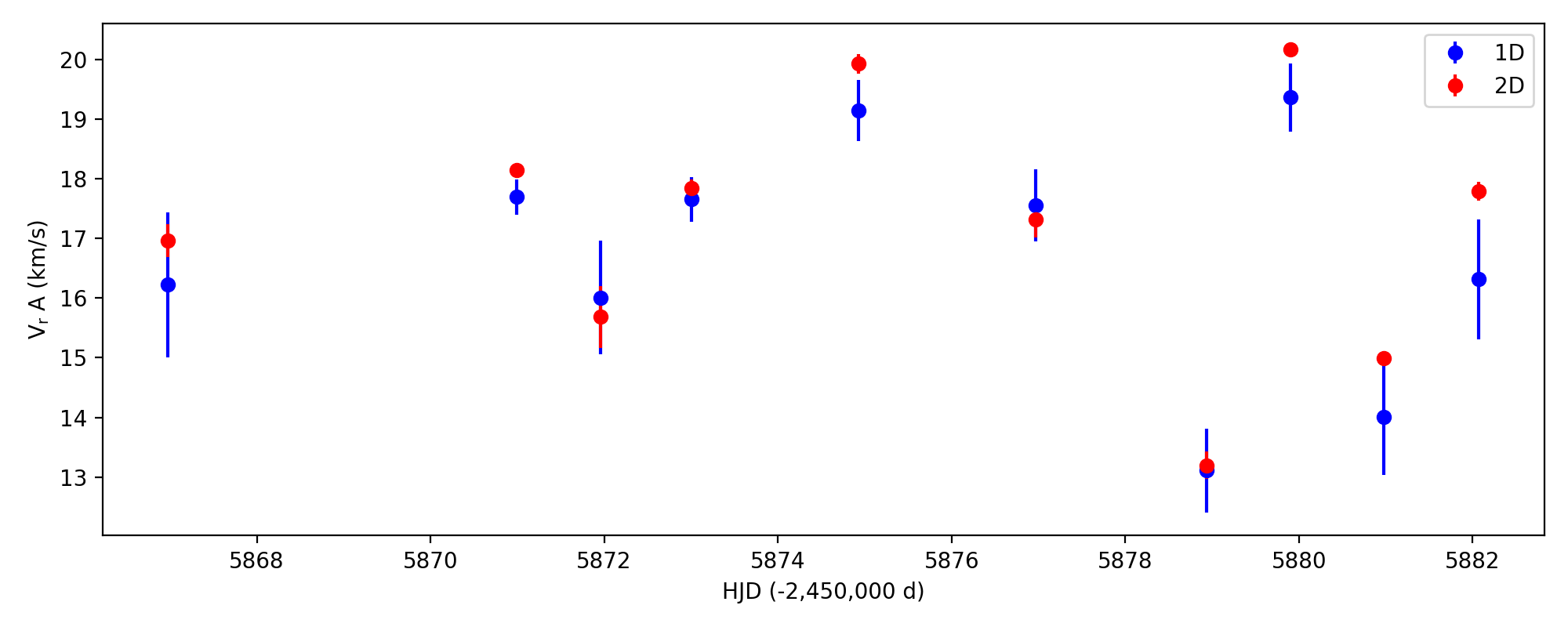}
        \caption{Radial velocity curves obtained using the primary only (\textit{blue}) and the primary plus secondary (\textit{red}) models.}
        \label{fig:comp1d2d}
    \end{figure}

\end{onecolumn}
    \newpage
    \section{Raw LSD profiles}
    \label{ap:rawLSD}

\begin{figure}[!h]
        \centering
        \includegraphics[width=0.49\linewidth]{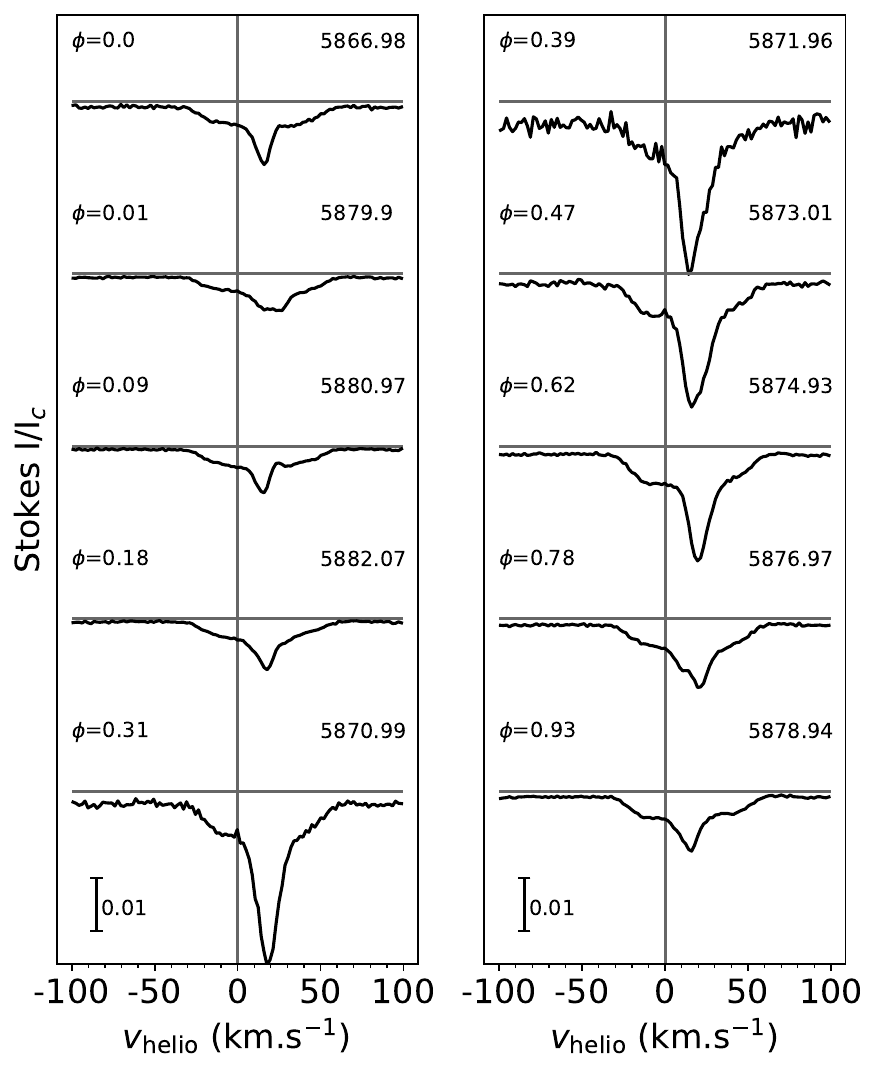}
        \includegraphics[width=0.49\linewidth]{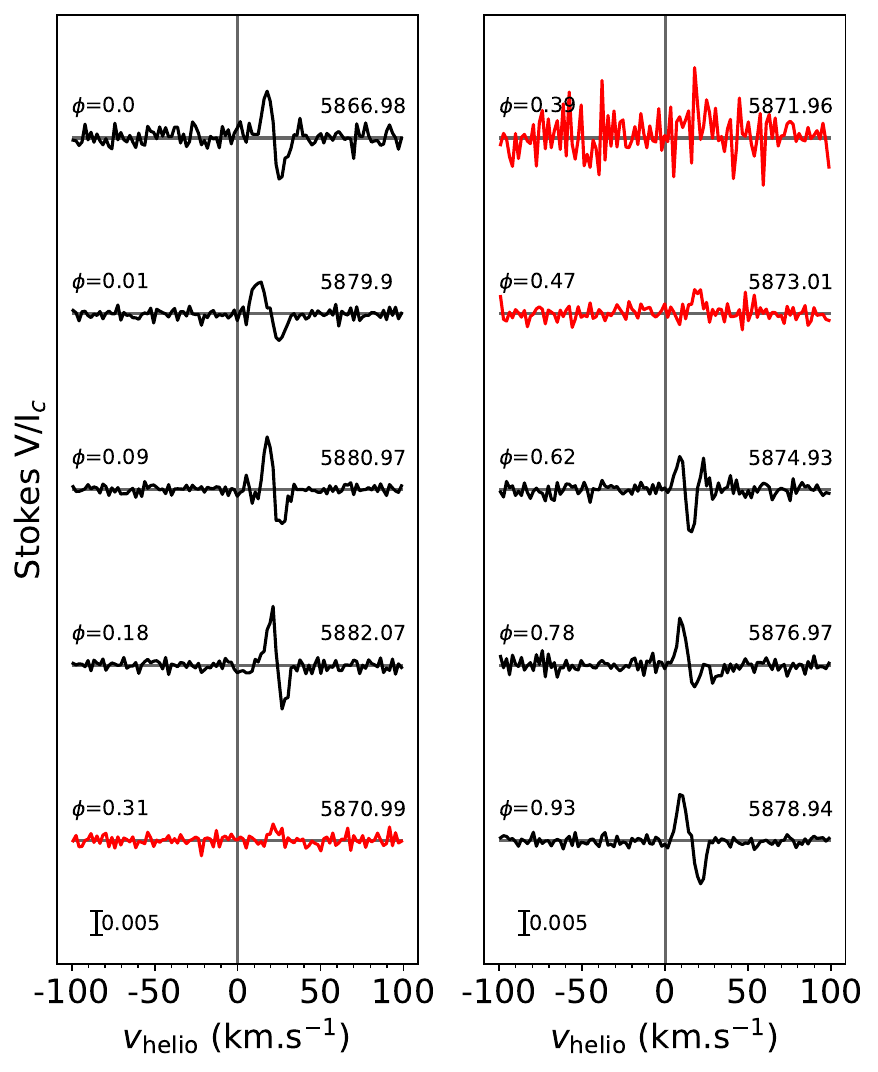}
        \caption{LSD Stokes \textit{I} (\textit{two left columns}) and Stokes \textit{V} (\textit{two right columns}) profiles of DF Tau. The rotation phases, computed using \prot=12.8 d (see Sect.~\ref{subsec:mag}), and the HJDs (-2\,450\,000 d) are indicated on the left and right of each profile, respectively. The profiles in red were divided by 3 for readability.}
        \label{fig:rawlsd}
    \end{figure}

\begin{figure}
        \centering
        \includegraphics[width=0.70\linewidth]{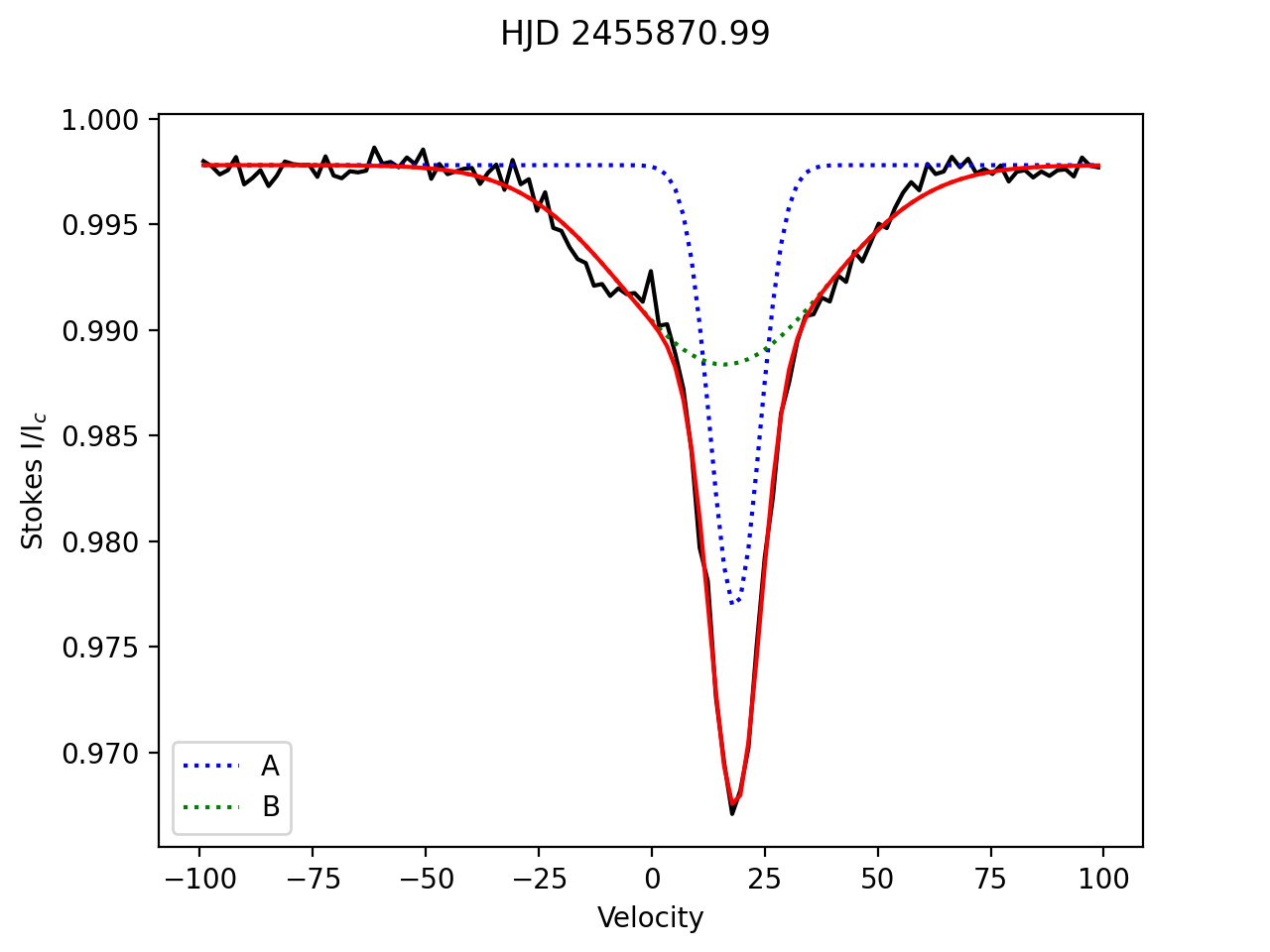}
        \caption{Example of the decomposition of the LDS profile at HJD 2\,455\,870.99. The observation is n black, the dotted blue and green lines are the A and B components, respectively. The red line is the total fit.}
        \label{fig:decompLSD}
    \end{figure}

    In this appendix we present the LSD profiles computed in Sect.~\ref{subsec:lsd}. These are shown in Fig.~\ref{fig:rawlsd} prior to the correction of the secondary's contribution.
    An example of the decomposition of the two components is shown in Fig.~\ref{fig:decompLSD}. 

\end{appendix}

\end{document}